\documentclass[sigconf,nonacm]{acmart}

\newcommand{\ProjectURL}{https://huggingface.co/collections/HPAI-BSC/ai-for-compilers-iris-family}
\newcommand{\supercomputer}{\texttt{MareNostrum 5}}
\newcommand{\av}[1]{\textcolor{blue}{\em #1 }}

\usepackage{listings}
\usepackage{subcaption}
\usepackage{pifont}
\usepackage{enumitem}

\usepackage{threeparttable}
\usepackage{pifont}

\usepackage{todonotes}

\AtBeginDocument{%
  }

\setcopyright{acmlicensed}
\copyrightyear{2018}
\acmYear{2018}
\acmDOI{XXXXXXX.XXXXXXX}
\acmConference[Conference acronym 'XX]{Make sure to enter the correct
  conference title from your rights confirmation email}{June 03--05,
  2018}{Woodstock, NY}
\acmISBN{978-1-4503-XXXX-X/2018/06}

\begin{document}

\title{LLM Translation of Compiler Intermediate Representation}

\author{Andrea Valenzuela Ramirez}
\email{andrea.valenzuela@bsc.es}
\orcid{0000-0002-2960-214X}
\affiliation{%
  \institution{Barcelona Supercomputing Center, Universitat Politècnica de Catalunya}
  \country{ }
}

\author{Cristian Gutierrez-Gomez}
\email{cristian.gutierrez@bsc.es}
\orcid{0009-0005-1441-8568}
\affiliation{%
  \institution{Barcelona Supercomputing Center}
  \country{ }
}

\author{Marta Barroso}
\email{marta.barroso@bsc.es}
\orcid{0000-0003-1396-6843}
\affiliation{%
  \institution{Barcelona Supercomputing Center}
  \country{ }
}

\author{Dario Garcia-Gasulla}
\email{dario.garcia@bsc.es}
\orcid{0000-0001-6732-5641}
\affiliation{%
  \institution{Barcelona Supercomputing Center}
  \country{ }
}

\author{Sara Royuela}
\email{sara.royuela@bsc.es}
\orcid{0000-0002-7644-0868}
\affiliation{%
  \institution{Barcelona Supercomputing Center}
  \country{ }
}

\renewcommand{\shortauthors}{Valenzuela et al.}

\begin{abstract}
    GCC and LLVM underpin much of modern software infrastructure, relying on distinct Intermediate Representations (IRs) to drive optimizations and code generation. However, the semantic and structural differences between these IRs create significant barriers for cross-toolchain interaction, limiting the reuse of compiler frontends, backends, and optimization pipelines across programming languages and compilation ecosystems. Traditional rule-based translators have attempted to bridge this gap, but their complexity and maintenance cost have hindered practical adoption. In this context, Large Language Models (LLMs) appear to be an emerging technology that offers a data-driven alternative, capable of learning complex mappings between heterogeneous compiler IRs directly from sufficiently representative examples. To explore this approach, this paper presents IRIS-14B, a 14-billion-parameter transformer model fine-tuned to translate GIMPLE (as emitted by GCC) to LLVM IR (as emitted by LLVM). The model is trained on paired IRs extracted from C sources and evaluated on the GIMPLE-to-LLVM IR transformation applied to IRs derived from real-world C code and competitive programming problems. To the best of our knowledge, IRIS-14B is the first model trained explicitly for IR-to-IR translation. It outperforms the accuracy of widely used models, including the largest state-of-the-art open models available today, ranging from 13 to 1,000 billion parameters, by up to 44 percentage points. The proposed transformation supports the integration of LLMs as complementary components within hybrid neuro-symbolic compiler architectures, where models such as IRIS-14B act as interoperability layers enabling cross-toolchain workflows without modifying existing compiler passes, while traditional compiler infrastructure continues to perform deterministic compilation and optimization.
\end{abstract}




\maketitle

\section{Introduction}

Compilers form the backbone of modern software infrastructure, and GCC and LLVM stand as the two most influential and widely deployed open-source compiler ecosystems. While GCC remains dominant in embedded and critical domains, LLVM has become predominant in emerging languages, hardware accelerators, and machine learning frameworks. Each system relies on its own Intermediate Representation (IR) stack, i.e., GENERIC, GIMPLE, and RTL in GCC, and LLVM IR and MIR in LLVM, to drive analyses, optimizations, and backend code generation. Among these, GIMPLE and LLVM IR occupy the middle-end of the compilation, preserving rich semantic information while remaining language- and target-independent, making them particularly suitable for optimizations.

The specialization of each compiler infrastructure and the normalization of heterogeneous architectures highlight the value of techniques that automatically translate between GIMPLE and LLVM IR. Such interoperability would enable practical scenarios for which no production-quality tools currently exist, empowering the community to combine strengths across ecosystems and build more modular, interoperable compilation pipelines. In particular, it would allow developers to exploit compiler-specific features unique to each system, including (a) GCC-specific extensions of common languages such as C and Fortran, e.g., nested functions (functions defined inside another function) and statically initialized flexible array members in C, and coarrays and asynchronous I/O operations in Fortran; (b) specific frontends, e.g., the Rust language is richly supported in LLVM, while legacy Modula-2 codebases are primarily supported via GCC-based frontends; (c) specific backends, e.g., most embedded targets are only supported in GCC; (d) mature optimizations for cross-polination workflows, e.g., domain specific optimizations from MLIR~\cite{mlir}/LLVM and software pipelining from GCC; and (e) sharing compiler-specific tooling, e.g., using LLVM Alive2~\cite{lopes2021alive2} in projects relying on GCC. An IR-to-IR translator would enable optimization analysis and verification, optimization cross-pollination, and the integration of GCC's and LLVM's frontends and backends.

Despite serving a similar role in the compilation pipeline, GIMPLE and LLVM IR differ substantially in their design. These differences include (a) \textit{granularity}, with GIMPLE being statement-oriented and LLVM IR expression-oriented; (b) \textit{structural model}, where GIMPLE preserves high-level control structures while LLVM IR lowers them into branches, $\varphi$ nodes, and basic blocks; (c) \textit{memory model}, with GIMPLE relying on virtual operands and LLVM IR using explicit load/store operations; (d) \textit{type system}, where GIMPLE is less strict and includes extensions; and (e) \textit{exception-handling}, which is abstracted in GIMPLE but explicit in LLVM IR.
These fundamental differences make the translation between the two IRs non-trivial, particularly when semantic preservation and robustness across real-world programs are required.

Even though the IRs differ in essence, the community has made efforts to bridge these two representations to unlock the reuse of compiler optimizations and the integration of new languages and hardware targets without duplicating substantial engineering effort. The original LLVM frontend, llvm-gcc~\cite{llvm04frontend}, extended GCC 3.x to generate LLVM IR and remained available from LLVM 1.3 until LLVM 2.9 (2011). Its successor, DragonEgg~\cite{gcc25dragonegg}, leveraged GCC's plugin infrastructure to improve modularity and was available from GCC 4.5/LLVM 2.7 until GCC 4.7/LLVM 3.3 (2013). Both tools relied on manually engineered translation passes and were instrumental during LLVM's early years, enabling access to mature GCC frontends and cross-toolchain experimentation. However, as both ecosystems evolved, the limitations of rule-based translators to generalize beyond narrow program subsets proved costly to maintain, and the consolidation of Clang led to the eventual deprecation of DragonEgg. More recently, Wyrm~\cite{rifkin2024wyrm}, a research GIMPLE-to-LLVM IR transpiler, and mlir-gccjit~\cite{mu2024mlirgccjit}, an MLIR dialect for libgccjit (an API for embedding GCC inside programs and libraries), illustrate the ongoing interest in connecting the two frameworks. However, they either remain exploratory or incomplete.

In contrast to traditional rule-based and pattern-matching solutions, Large Language Models (LLMs) offer a data-driven alternative that can infer structural correspondences and learn semantics from diverse corpora, making them a promising foundation for accurate IR-to-IR translation. Models such as GPT~\cite{achiam2023gpt}, DeepSeek~\cite{guo2025deepseek}, and Qwen~\cite{bai2023qwen} demonstrate strong statistical understanding of source code~\cite{nam2024using,pan2025code}. Yet relatively little emphasis has been placed on training models directly on the compiler IR, even though IRs expose structural and semantic properties that can support tasks such as binary decompilation~\cite{toor2022decompilation}, and code lifting~\cite{tan2023splendid}.

The role of IRs in current LLM training pipelines remains underexplored. Standard code datasets primarily contain high-level source code scraped from online repositories, yet source code can easily be compiled down to IR and assembly. This observation enables richer training setups in which paired high-level and low-level representations coexist. Furthermore, compiling programs across multiple optimization levels and target architectures naturally produces aligned IR sets that preserve semantics while exposing diverse structural transformations. Such datasets are promising resources for training LLMs to learn IR-to-IR mappings that would be extremely difficult to encode with handcrafted rules.

This work introduces IRIS-14B, an open-source transformer model specifically trained for GIMPLE-to-LLVM IR translation. To the best of our knowledge, IRIS-14B is the first LLM specifically trained for IR translation. The model is evaluated in terms of syntactic correctness and semantic equivalence, and compared against state-of-the-art open models. To better understand the model's capabilities, we further study the properties of the codes the model successfully translates and compare them with those it fails to translate. In addition, we conduct experiments to characterize the nature of code samples that achieve better accuracy when training LLM-based IR-to-IR translators. Finally, we present two use cases that demonstrate the model's ability to generalize to previously unseen code while preserving program semantics. Overall, this paper makes the following contributions:
\begin{enumerate}
    \item A novel methodology for performing IR-to-IR translation using LLMs integrated with the existing compiler toolchains.
    \item IRIS-14B, an open-source 14-billion-parameter model fine-tuned for GIMPLE-to-LLVM IR translation. The model achieves state-of-the-art correctness on two representative IR-to-IR evaluation benchmarks, substantially outperforming general-purpose code models of comparable and larger size up to 44 percentage points over the strongest baseline.
    \item A suite of datasets specially tailored for training and evaluation. The training data comprise two datasets of paired GIMPLE and LLVM IR, \textit{TheStack-IRIS} and \textit{GNU-IRIS}, derived from \textit{TheStack} and GNU utilities code corpora, respectively. The evaluation sets are based on the existing \textit{ExeBench} and \textit{CodeForces} datasets, which we adapt to the IR translation task as \textit{ExeBench-IRIS} and \textit{CodeForces-IRIS}.
    \item An evaluation pipeline based on syntactic correctness and functional equivalence assessed through I/O tests for correctness verification, together with an analysis of model failure modes and a study of the characteristics of code that lead to higher IR-to-IR translation accuracy.
    \item A through discussion about the implications of extending IRIS-14B to different programming languages, the role of LLM-based IR-to-IR translation in future compilation pipelines, including the evolution of compiler versions, and the LLVM IR-to-GIMPLE translation direction.
\end{enumerate}

These contributions form a solid proof of concept for future research on leveraging LLMs to improve interoperability between the LLVM and GCC compiler infrastructures, providing a practical and extensible tool for both academic and industrial communities to accelerate the adoption of new programming languages and hardware platforms and to stimulate optimization research, thereby enabling techniques such as optimization cross-pollination.
\section{Domain context}

\begin{figure*}[h!]
    \centering
    \includegraphics[width=0.75\linewidth]{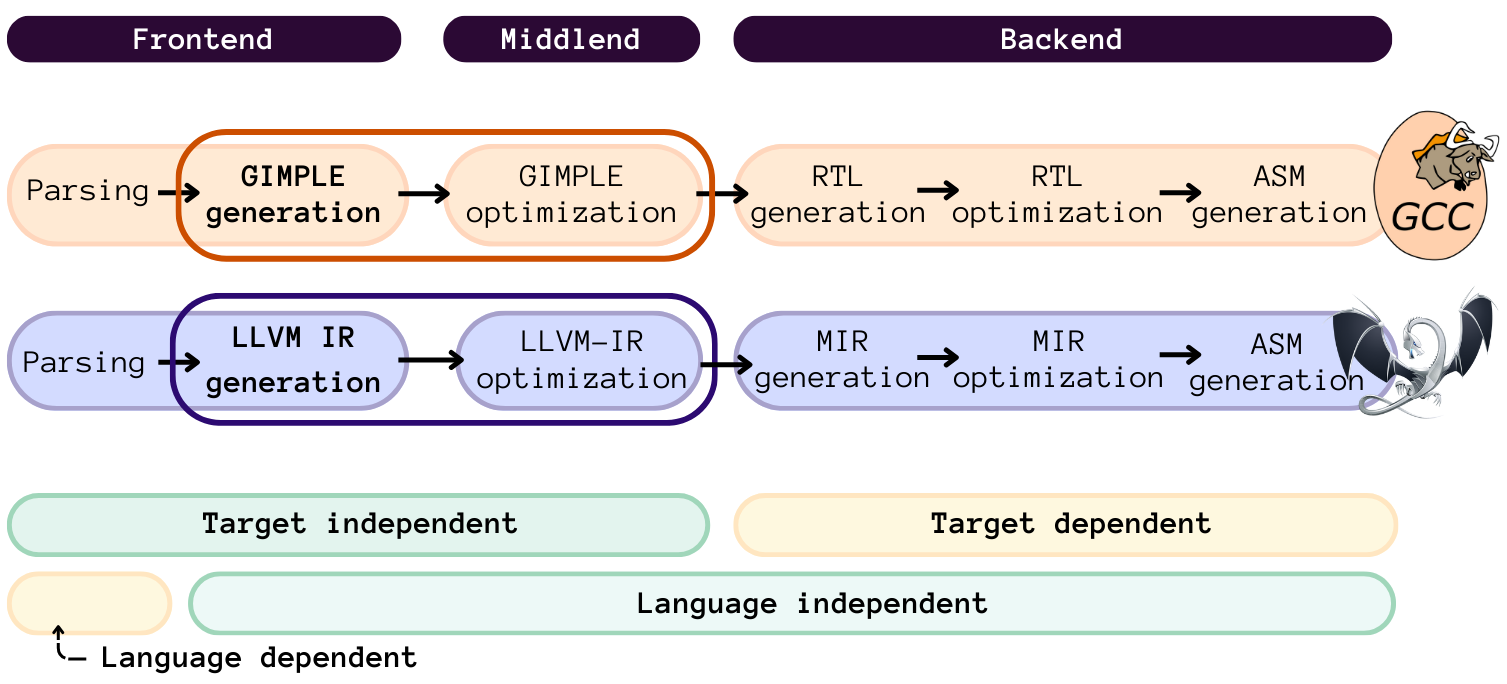}
    \caption{GCC and LLVM generic compilation pipelines, highlighting steps using GIMPLE and LLVM IR, respectively.}
    \label{fig:compiler-pipelines}
\end{figure*}

Compilers translate programs written in high-level languages into executable code for a target machine. They are typically organized into three parts: (i) a \textit{frontend}, which parses the source program, builds an Abstract Syntax Tree (AST) representing the hierarchical structure of the code, and lowers the code into an IR that captures the semantics of the program in a somewhat language-independent form; (ii) the \textit{middle-end}, which performs analysis and language- and target-agnostic optimizations; and (iii) the \textit{backend}, which transforms the IR into instructions tailored to the target architecture, while performing target-aware optimizations.

IRs play a central role in compilation pipelines because they decouple the language-specific concerns handled by frontends from the architecture-specific concerns handled by backends. By operating on IR, compiler optimizations can be reused across multiple programming languages and hardware targets. For this reason, modern compiler infrastructures typically employ several IRs with different abstraction levels, ranging from high-level representations that retain structural information from the source program to lower-level representations closer to machine instructions.

For decades, GCC (GNU Compiler Collection) held a hegemonic position in the open-source compiler landscape. Its monolithic design integrates frontends for multiple languages, including C/C++, Fortran, Modula-2, Ada, and Go, with tightly coupled middle- and back-ends. While extremely powerful, the interleaved optimizations and limited modular interfaces make extending GCC to new languages or architectures, as well as maintenance and experimentation, challenging.

LLVM emerged in 2003 as a modular and extensible alternative compiler infrastructure. LLVM supports a broad set of languages today through frontends such as Clang for C and C++, Flang for Fortran, and community compilers for Rust, Swift, Julia, D, and Haskell. Its clear interfaces and modular pipeline ease maintenance and facilitate evolution, which has contributed to its rapid adoption in domains such as emerging languages, accelerators, and machine learning. However, LLVM remains less prevalent in safety-critical and embedded systems, and legacy and specialized hardware, where GCC has historically dominated.

The two compilers share a similar pipeline, both depicted in Figure~\ref{fig:compiler-pipelines}. GCC lowers source languages into GIMPLE, either directly (as in C, C++) or passing through GENERIC first (as in Fortran, Ada, Go), using a process called \textit{gimplification}. GIMPLE is a language- and target-independent IR on top of which GCC performs high-level optimizations like constant propagation and loop transformations. Then, it generates RTL to perform target-dependent optimizations, such as register allocation and instruction pipelining, before generating assembly code. LLVM follows a similar approach: the frontend generates LLVM IR, a language-agnostic, target-independent IR on top of which high-level (e.g., dead code elimination) as well as low-level (e.g., vectorization) optimizations are performed. LLVM IR is later lowered to MIR, a low-level target-specific representation where architecture-specific optimizations are performed. 

Although GCC and LLVM have similar pipelines, their IRs differ substantially. Figure~\ref{lst:compiler-irs} illustrates these differences by showing a C code snippet that adds two numbers (Figure~\ref{lst:c}), and the corresponding GIMPLE (Figure~\ref{lst:gimple}) and LLVM IRs (Figure~\ref{lst:llvm-ir}), both generated without optimizations. GIMPLE provides a relatively high-level abstraction that remains close to the structure of the source code, that expresses computation as a sequence of simple operations, where temporary variables are introduced to store intermediate values, statements are written in a three-address form, and often Static Single Assignment (SSA) form, and structured control constructs, including loops and conditionals, are lowered to explicit conditional and unconditional jumps. Furthermore, GIMPLE often reflects language-specific features originating from differences in frontend lowering strategies, type systems, and runtime libraries. LLVM IR, on the other hand, provides a lower-level, more uniform representation based on SSA. Program structure is represented through basic blocks connected by branches and $\varphi$ nodes, and memory operations are expressed explicitly through \texttt{load} and \texttt{store} instructions. LLVM also enforces a stricter type system and a more explicit memory model than GIMPLE, resulting in a representation that is more regular but further away from the original program's high-level semantics.

\begin{figure*}[t]
  \centering

  \begin{minipage}[t]{0.47\textwidth}
    \centering

    \begin{subfigure}[t]{\linewidth}
      \begin{lstlisting}[language=C, basicstyle=\ttfamily\small]
int add(int a, int b) {
  return a + b;
}

int main() {
  int x = add(2, 3);
  return x;
}
      \end{lstlisting}
      \vspace{-0.25cm}
      \caption{C source code.}
      \label{lst:c}
    \end{subfigure}

    \vspace{0.15cm}

    \begin{subfigure}[t]{\linewidth}
      \begin{lstlisting}[language=C, basicstyle=\ttfamily\small]
int __GIMPLE (int a, int b) {
  int D_2841;
  D_2841 = a + b;
  return D_2841;
}

int __GIMPLE () {
  int D_2843;
  {
    int x;
    x = add (2, 3);
    D_2843 = x;
    return D_2843;
  }
  D_2843 = 0;
  return D_2843;
}
      \end{lstlisting}
      \vspace{-0.25cm}
      \caption{GIMPLE IR \mbox{(-fdump-tree-gimple)}.}
      \label{lst:gimple}
    \end{subfigure}

  \end{minipage}
  \hfill
  \begin{minipage}[t]{0.50\textwidth}
    \centering

    \begin{subfigure}[t]{\linewidth}
      \begin{lstlisting}[language=LLVM, basicstyle=\ttfamily\small]
define dso_local i32 @add(i32 noundef %a,
                          i32 noundef %b) {
entry:
  %a.addr = alloca i32, align 4
  %b.addr = alloca i32, align 4
  store i32 %a, ptr %a.addr, align 4
  store i32 %b, ptr %b.addr, align 4
  %0 = load i32, ptr %a.addr, align 4
  %1 = load i32, ptr %b.addr, align 4
  %add = add nsw i32 %0, %1
  ret i32 %add
}

define dso_local i32 @main() {
entry:
  %retval = alloca i32, align 4
  %x = alloca i32, align 4
  store i32 0, ptr %retval, align 4
  %call = call i32 @add(i32 noundef 2,
                        i32 noundef 3)
  store i32 %call, ptr %x, align 4
  %0 = load i32, ptr %x, align 4
  ret i32 %0
}
      \end{lstlisting}
      \vspace{-0.25cm}
      \caption{LLVM IR \mbox{(-S -emit-llvm)}.}
      \label{lst:llvm-ir}
    \end{subfigure}

  \end{minipage}

  \caption{Different representations of a code snippet adding two integers. Compilers: LLVM 21.1 for LLVM IR and GCC 15.2 for GIMPLE, all using \mbox{-O0}.}
  \label{lst:compiler-irs}
\end{figure*}

The semantic mismatches between GIMPLE and LLVM IR pose challenges to handcrafted rule-based mapping approaches, such as DragonEgg~\cite{gcc25dragonegg}, and more recent experimental tools, such as Wyrm~\cite{rifkin2024wyrm}. DragonEgg, a GCC plugin that replaced GCC’s optimizers and code generators with LLVM’s, generated LLVM IR directly from GIMPLE. Its support, however, was limited to a subset of languages (C, C++, Fortran) and to specific GCC/LLVM version combinations, and it remains deprecated since GCC 4.7/LLVM 3.3 (2013). Translating structured control flow, exception handling, and GCC-specific extensions such as statically initialized flexible array members and nested functions proved particularly difficult. Over time, the project became tied to older compiler versions, making continued maintenance impractical as toolchains evolved. Similarly, contemporary projects like Wyrm focus on GIMPLE-to-LLVM IR translation but remain experimental and incomplete.

Both DragonEgg and Wyrm struggle because the two IRs embody different design philosophies: GIMPLE reflects GCC’s internal, structured, and often implicit semantics, while LLVM IR has a stricter, more explicit instruction and type model, making features such as implicit type conversions, calling conventions, and GCC‑specific extensions complex to translate soundly. Consequently, translating between the two IRs requires recovering and restructuring semantic information rather than applying simple syntactic rewrites. In practice, any evolution in GIMPLE, GCC frontend extensions, or LLVM’s IR semantics forces corresponding changes in translation rules, which are brittle and costly to maintain, and motivates exploring alternative approaches that automatically learn semantic correspondences.

\section{Related work}

The task of translating GIMPLE-to-LLVM IR is currently addressed only by the Wyrm~\cite{rifkin2024wyrm} experimental project. Based on the limited documentation available, Wyrm supports a subset of GIMPLE types, instructions, and operators, but lacks comprehensive coverage of features such as target calling conventions. In addition, Wyrm relies on the experimental GIMPLE-Frontend to process GIMPLE inputs and therefore does not directly accept raw GIMPLE dumps produced by standard compiler passes. As a result, even in the GIMPLE-to-LLVM direction, Wyrm requires manual post-processing of the input before translation. Moreover, its public repository has not been updated since its initial release in 2024, suggesting that the approach faces maintainability and scalability challenges. Wyrm reinforces two broader observations: (a) GIMPLE-to-LLVM IR is a relevant task, and (b) manually engineered IR-to-IR translators struggle to keep pace with evolving compiler infrastructures and complex missmatches.

In parallel, advances in machine learning have reshaped the landscape of language processing. Transformers~\cite{vaswani2017attention} have emerged as the dominant architecture for natural language processing (NLP), enabling models to capture long-range dependencies and contextual information. Large transformer-based models such as GPT~\cite{achiam2023gpt} and Qwen~\cite{bai2023qwen} have demonstrated remarkable capabilities across a variety of NLP tasks, including text classification, translation, and generation.

Building on these successes, researchers have extended transformer models beyond natural language to programming languages, leveraging the structural regularities and formal semantics of code. AI systems such as GitHub Copilot~\cite{github25copilot}, OpenAI Codex~\cite{openai25codex}, and DeepMind AlphaCode~\cite{li2022competition} show that large-scale transformers can model program syntax and semantics sufficiently well to assist with code generation and completion, thereby increasing developers’ efficiency \cite{yeticstiren2023evaluating}, though not without controversy.

More recently, transformers have been heavily applied to source-to-source code translation, or \textit{transpilation}, between high-level programming languages \cite{roziere2020unsupervised,eniser2024towards,valenzuela2025from,ranasinghe2025llm}. While promising, source-level translation remains challenging due to semantic drift across syntactically dissimilar languages, the large context requirements of real-world programs, and the difficulty of ensuring deterministic, semantics-preserving outputs for functionally equivalent code. 

Extending transformer applications to compiler IRs addresses many limitations of source-level code translation. Unlike high-level languages, IRs like GIMPLE and LLVM IR encode detailed semantics, control flow, and type information in a structured, language-agnostic manner. Prior work has shown that transformers can learn to translate between C source code and LLVM IR~\cite{moses2022understanding}, indicating that these models can capture the semantics required for IR-level transformations. Jiang et al.~\cite{jiang2025can} evaluated the ability of popular LLMs to understand and manipulate compiler IRs, showing that general-purpose models can parse IR syntax and recognize high-level structures but consistently struggle with control-flow reasoning, execution semantics, and loop handling. Further related work has leveraged low-level compiler IRs to improve the accuracy of code translation tasks~\cite{szafraniec2022code,paul2024ircoder}.

While recent work suggests that LLMs can, in principle, operate at the level of compiler IRs, their effectiveness is fundamentally constrained by the availability of suitable training data. Existing curated datasets are overwhelmingly centered on high-level programming languages, such as Python and Java, with compiler IRs representing only a marginal fraction of the collected data. Some prior efforts employ IRs as an intermediate representation for high-level language translation, releasing their data ~\cite{szafraniec2022code}, while others build corpora that pair source programs with their corresponding LLVM IR~\cite{cummins2023large,grossman2023compile}. However, these resources focus exclusively on a single compiler ecosystem (LLVM) and do not provide aligned IR corpora spanning multiple compilers, leaving cross-IR translation largely unexplored.

This work presents IRIS-14B, a model fine-tuned for GIMPLE-to-LLVM IR translation, along with a novel methodology that integrates IRIS into GCC and LLVM compiler infrastructures, offering a complete compilation pipeline. To our knowledge, IRIS-14B is the first LLM designed specifically for IR-to-IR translation between heterogeneous compiler ecosystems, a task that has historically relied on rule-based approaches. We also address the data gap by releasing the first publicly available datasets of semantically equivalent GIMPLE–LLVM IR extracted from C programs. These datasets can be used either as standalone IR corpora or as aligned pairs for fine-tuning and evaluating models on IR-to-IR translation, IR optimization, and other low-level code tasks.
\section{Methodology}

This section presents the methodology used to develop IRIS-14B. Subsection~\ref{sec:translation-method} describes the IRIS translation methodology, namely how IRIS-14B bridges the GCC and LLVM compiler ecosystems. Subsection~\ref{sec:training} details the training methodology used to build the model, and Subsection~\ref{sec:eval} presents the evaluation methodology used to assess its accuracy in the GIMPLE-to-LLVM IR translation task.

\subsection{Translation methodology}\label{sec:translation-method}

\begin{figure}[H]
    \centering
\includegraphics[width=0.99\linewidth]{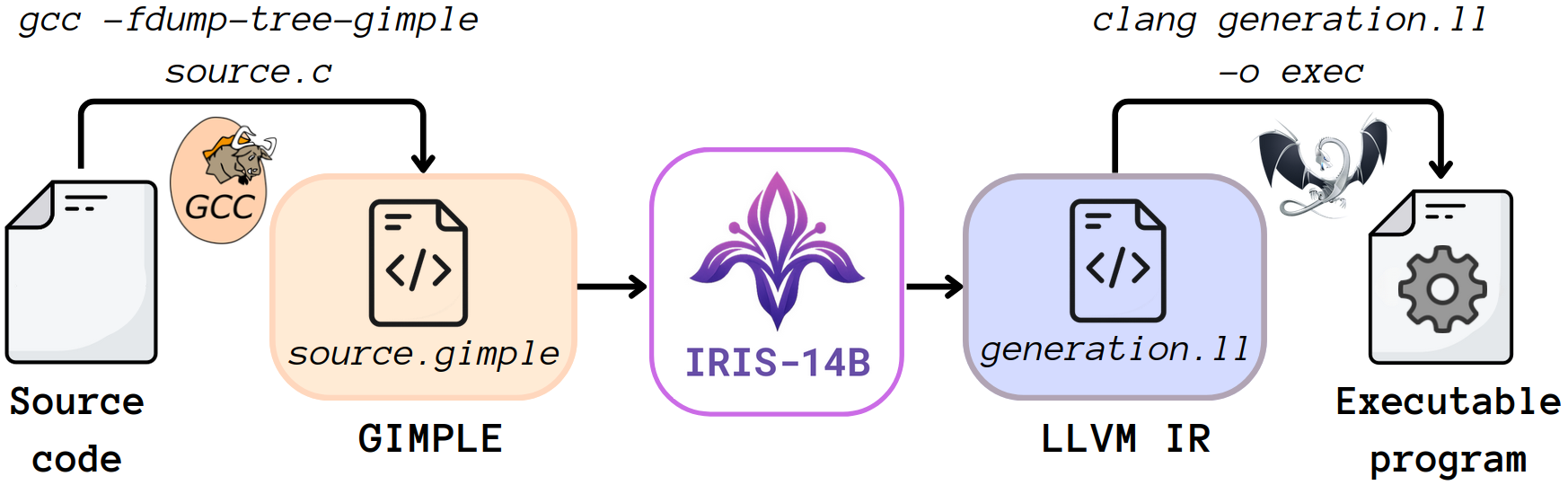}
    \caption{IRIS methodology for the integration of GCC and LLVM through GIMPLE-to-LLVM IR translation.}
    \label{fig:iris-workflow}
\end{figure}

The overarching goal of this work is to integrate the GCC and LLVM compiler infrastructures for compiling a given source code by automatically translating from GIMPLE to LLVM IR. Figure~\ref{fig:iris-workflow} illustrates the IR-to-IR methodology proposed, which is composed of the following steps: (1) generate the unoptimized GIMPLE representation of a given source code, (2) use the IRIS model for GIMPLE-to-LLVM IR translation, and (3) compile the resulting LLVM IR with LLVM for assembly generation. In this context, we define the IRIS task as the translation of GIMPLE into LLVM IR. This task is formalized as follows:
\begin{description}[leftmargin=!,labelwidth=\widthof{\bfseries Instruction:}]
    \item [Input:] Unoptimized GIMPLE, as emitted by the \texttt{-fdump-tree}\allowbreak\texttt{-gimple} flag of GCC.
    \item [Instruction:] Translate the input into its LLVM IR counterpart.
    \item [Goal:] The \texttt{.ll} file containing the generated LLVM IR enables LLVM compilation to produce the corresponding executable.
\end{description}

Note that the compiler toolchains do not require any additional modifications. The IRIS methodology has been intentionally designed to be decoupled from compiler internals, operating on textual IR at both input and output. As a result, only changes in IR syntax would require updating the model. This avoids modifications to either toolchain and reduces maintenance as the toolchains evolve, one of the main reasons behind DragonEgg’s discontinuation.

\subsection{IRIS training} \label{sec:training}

\begin{figure}[H]
    \centering
\includegraphics[width=0.99\columnwidth]{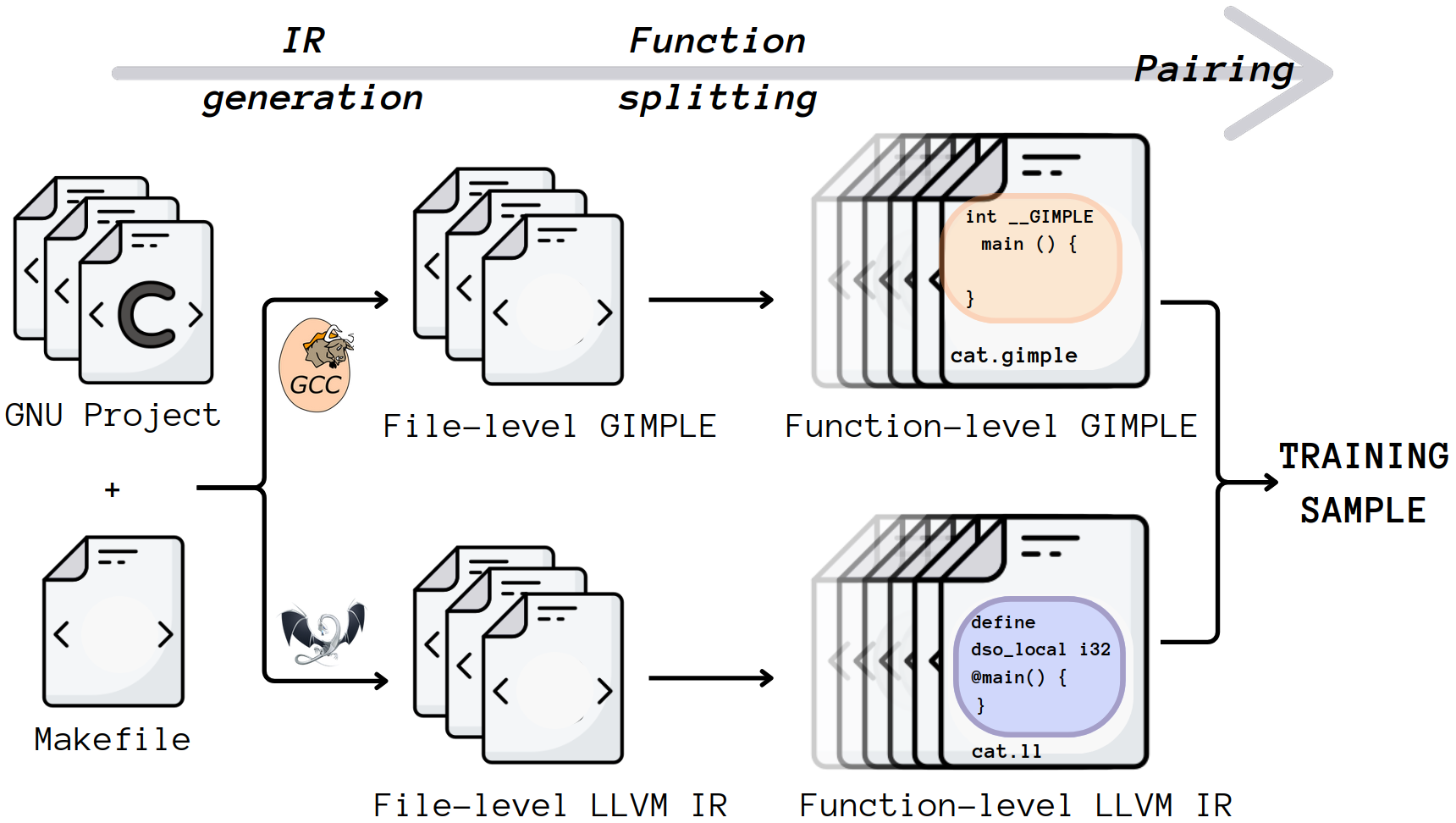}
    \caption{Diagram of the proposed approach to extract IR pairs of function-level real-world code.} 
    \label{fig:iris-gnu}
\end{figure}

This work uses two sources of data to generate pairs of GIMPLE and LLVM IR representations for training: \textit{TheStack} dataset~\cite{kocetkov2022stack} and \textit{GNU} code repositories~\cite{gnu_packages}. The former, \textit{TheStack}, comprises source code from GitHub repositories across 30 programming languages, from which only C code is selected and deduplicated. Crafting the IR version of this dataset requires compiling the samples into an object, so code snippets that do not compile are also filtered out, yielding around 310K C code samples from which we extract both GIMPLE and LLVM IR. The resulting paired IR version of \textit{TheStack} is released under the name \textit{TheStack-IRIS}. The latter, released as \textit{GNU-IRIS}, is a dataset built for this work, comprising 13,049 aligned function pairs from selected \textit{GNU utils} repositories. Figure~\ref{fig:iris-gnu} illustrates the approach to generating training data. While for simple applications the IR can be dumped at the source file level to be used directly for training, for some real-world repositories such as the ones from \textit{GNU}, file-level IR dumps become very long and can easily exceed the maximum input context length supported by current models (see Section~\ref{sec:context} for further discussion of the impact of context length). To use large repository-level code while staying within context limits, we process file-level IR into function-level samples as follows: (1) modify the build configuration to inject flags that request GIMPLE and LLVM IR dumps during compilation; (2) parse these dumps to extract, for each C function, the corresponding GIMPLE and LLVM IR functions; (3) store each resulting triplet \texttt{\{C function, GIMPLE function, LLVM function\}}, using the GIMPLE–LLVM pairs as the actual training samples.

IRIS-14B builds on top of the transfer learning paradigm, which uses a strong pre-trained model as a starting point to maximize model quality. \textit{Qwen3}~\cite{yang2025qwen3}, a recent small-to-medium-scale model with 14B parameters, is selected as the source model due to its manageable size. Qwen3 is based on a decoder-only transformer architecture trained autoregressively to generate sequences by iteratively predicting the next token conditioned on the prior context. This design makes it well-suited for modeling long-range dependencies, such as those found in code. On top of this pre-trained model, fine-tuning is performed on 1.4B tokens of paired training data derived from the \textit{TheStack-IRIS} and \textit{GNU-IRIS} datasets as defined before and summarized in Table \ref{tab:datasets-reference}. Training is conducted using the \textit{AdamW} optimizer~\cite{loshchilov2017decoupled} with $\beta_1$ and $\beta_2$ values of $0.9$ and $0.999$. A cosine scheduler with 3\% warm-up ratio is employed, with the peak learning rate set to $2.07e^{-5}$. The maximum sequence length during training is $2^{14}$ tokens, and the full training procedure spans three epochs over the corpus. The model resulting from this training procedure, \textbf{IRIS-14B}~\footnote{IRIS-14B model: \url{https://huggingface.co/HPAI-BSC/IRIS-14B}}, is openly released with this work, together with the training datasets \textit{\textit{TheStack-IRIS}}~\footnote{TheStack-IRIS: \url{https://huggingface.co/datasets/HPAI-BSC/TheStack-IRIS}} and \textit{GNU-IRIS}~\footnote{GNU-IRIS: \url{https://huggingface.co/datasets/HPAI-BSC/GNU-IRIS}}.
IRIS-14B has been trained on the $\supercomputer$ using compute nodes with 4xH100 NVIDIA GPUs. The training completes in approximately 35 hours on 15 nodes, consuming 0.86 MWh of energy, which corresponds to an estimated 244 kg of $CO_2$ emissions.

\subsection{IRIS evaluation} \label{sec:eval}

The evaluation of the model accuracy on the IRIS task (i.e., GIMPLE-to-LLVM IR translation) is performed on two data sources distinct from and disjoint from those used during training: \textit{ExeBench} \cite{armengol2022exebench} and \textit{CodeForces} \cite{penedo2025codeforces}.

\textit{ExeBench} contains a collection of executable C functions extracted from real code repositories, providing (1) a C++ wrapper for each sample that allows the code to be compiled and executed, and (2) input/output (I/O) pairs for verifying functional equivalence. \textit{ExeBench} includes two types of samples: \emph{real} samples, where the original auxiliary definitions (header files and external functions and types) are recovered from the corresponding GitHub repository, and \emph{synthetic} samples, where dependencies are generated synthetically. The evaluation is performed on a selection of the \emph{real} samples, which contains 2,134 snippets, discarding samples that do not compile with both GCC and Clang compilers, or that fail any of the I/O tests. 

While \textit{ExeBench} preserves the characteristics of real-world code, the authors apply extensive post-processing to extract standalone, executable functions and to generate input–output test cases, ensuring that samples remain manageable in size and sufficiently diverse in structure. As a result, \textit{ExeBench} enables evaluation scenarios that resemble competitive programming benchmarks in terms of self-contained executability and test-driven validation, while preserving real-world code characteristics.

A practical challenge of using \textit{ExeBench} for IR-to-IR translation is compiling the model-generated IR with the C++ wrapper of each sample (required by the benchmark for sample execution). To achieve this, we propose the following pipeline: (1) the model-generated LLVM IR is compiled to an object file with \texttt{llc}; (2) all required C declarations, extracted with \texttt{ctags}, are added to the wrapper as \texttt{extern "C" \{\}} declarations; and (3) the C++ wrapper is linked with the object file produced from the model-generated IR with \texttt{clang++}. Before evaluation, the ground-truth LLVM IR is verified to support the proposed workflow, discarding any samples that do not. After this filtering, 1,764 samples are retained for evaluation.

\textit{CodeForces} is a major online competitive programming platform that regularly hosts contests and maintains an extensive, publicly available archive of algorithmic problems and user submissions. We build another IR-to-IR evaluation set from this archive, selecting a total of 488 problems containing user submissions in C after validating those samples by compiling them with both GCC and Clang compilers. To ensure correctness and code quality, we further validate that submissions pass all the platform I/O tests within a time limit of 15 seconds per test, a reasonable time margin for a competitive programming domain, and as a filtering to include the top-performing submissions.

To reduce redundancy among submissions for the same \textit{CodeForces} problem, we represent each C submission as a feature vector of static and dynamic metrics extracted by parsing the implementation’s AST and executing the code. Static features capture program structure and complexity, while dynamic features expose runtime behavior. For static metrics, we consider 13 metrics across several categories: (i) \textit{variable-related} metrics, including the number of global mutable and global constant variables; (ii) \textit{control-flow} metrics, accounting for conditionals and loops; (iii) \textit{memory-related} operations, including the use memory management functions such as \texttt{malloc()}; (iv) \textit{complexity metrics}, including lines of code and nesting depth; (v) \textit{array metrics}, including the number of instantiated arrays and the number of array read and write accesses; (vi) \textit{pointer-related} metrics, covering the number of instantiated pointers (both \texttt{void} and \texttt{typed}), pointer calls, and pointer arithmetic operations; and (vii) \texttt{struct} usage. For dynamic metrics, we include wall-clock time, peak memory, CPU utilization, and executable size. 

For every problem, $k$-means clustering with $k=3$ is applied to group submissions into three clusters using the feature space presented above. From each cluster, the submission closest to the cluster centroid is selected as its representative. The working hypothesis is that each representative corresponds to a distinct implementation strategy for the underlying computation, maximizing code diversity while minimizing redundancy in the dataset. Finally, we compile all selected submissions across all problems to extract their corresponding GIMPLE representations. This yields 1,192 GIMPLE-to-LLVM IR translation tasks, which are used as evaluation samples.

Both evaluation datasets of IR pairs, curated and compiled from \textit{ExeBench} and \textit{CodeForces}, are openly released with this work as \textit{ExeBench-IRIS}~\footnote{ExeBench-IRIS: \url{https://huggingface.co/datasets/HPAI-BSC/ExeBench-IRIS}} and \textit{CodeForces-IRIS}~\footnote{CodeForces-IRIS: \url{https://huggingface.co/datasets/HPAI-BSC/CodeForces-IRIS}}, respectively. Table \ref{tab:datasets-reference} summarizes all datasets used in this work, indicating their size and their purpose in this work. All GIMPLE and LLVM IR textual representations used for training and evaluation are extracted using latest compiler versions, GCC-15.2.0 and Clang-22.1.0, respectively.

The following section uses both datasets to evaluate model accuracy in IR translation with respect to syntactic correctness and functional equivalence. GIMPLE lacks a formal specification, and although LLVM IR is documented, it does not provide complete formal semantics. As a result, semantic preservation cannot be established through formal equivalence proofs, a limitation that applies to both rule-based and learning-based translators. Instead, following established compiler engineering practice, syntactic correctness is assessed by checking whether the generated IR compiles, while functional equivalence is assessed by executing the corresponding I/O tests on the resulting binary. 

\begin{table}[t]
\centering
\caption{Summary of the datasets used in this work, indicating whether each dataset is used for training (Section~\ref{sec:training}) or evaluation (Section~\ref{sec:eval})}
\label{tab:datasets-reference}
\begin{threeparttable}
\begin{tabular}{lr|cc|}
\textbf{Dataset name} & \textbf{Size} & \textbf{Train} & \textbf{Test} \\
\midrule
\textit{TheStack-IRIS}    & 310k  & \ding{51} & \ding{55} \\
\textit{GNU-IRIS}         & 13k & \ding{51} & \ding{55} \\
\textit{CodeForces-IRIS}  & 1.2k  & \ding{51}\tnote{$^{1}$}  & \ding{51} \\
\textit{ExeBench-IRIS}    & 1.7k  & \ding{55} & \ding{51} \\
\end{tabular}
\begin{tablenotes}
\footnotesize
\item[1] Subset used for training only in Section~\ref{sec:code-types}.
\end{tablenotes}
\end{threeparttable}
\end{table}

\begin{table*}[h!]
\centering
\caption{\texttt{pass@1} with \texttt{N=3} for the IRIS task (GIMPLE-to-LLVM IR translation) of different LLMs on the two different test sets. Models ordered by number of parameters (model size), except for IRIS-14B, which is shown in the last row and highlighted in bold.}
\label{tab:model_performance}
\begin{tabular}{l@{}r|rrrr}
 & & \multicolumn{2}{c}{\textbf{CodeForces-IRIS}} & \multicolumn{2}{c}{\textbf{ExeBench-IRIS}} \\
\cmidrule(lr){3-4} \cmidrule(lr){5-6}
\textbf{Model} & \textbf{Params} & \textbf{Compile} (\%) & \textbf{I/O Test} (\%) & \textbf{Compile} (\%) & \textbf{I/O Test} (\%)\\
\toprule
Kimi-K2-Instruct-0905 \cite{team2025kimi} & 1000 B & 13.83 & 10.33 & 68.86 & 48.49 \\
Qwen3-Coder-A35B \cite{yang2025qwen3} & 480 B & 13.17 & 10.07 & 71.57 & 49.91 \\
gpt-oss-120b \cite{agarwal2025gpt} & 120 B & 23.43 & 19.30 & 80.54 & 72.84 \\
gpt-oss-20b \cite{agarwal2025gpt} & 20 B & 11.44 & 8.22 & 74.00 & 61.32 \\
Qwen3-14B \cite{yang2025qwen3} & 14 B & 0.06 & 0.02 & 42.39 & 31.76 \\
LLM-Compiler-13B \cite{cummins2025llm} & 13 B & 80.93 & 0.45 & 61.80 & 23.23 \\
\midrule
\textbf{IRIS-14B} &\textbf{ 14 B} & \textbf{73.24} &  \textbf{63.26} & \textbf{86.89} & \textbf{79.06} \\
\end{tabular}
\end{table*}

\section{Experimentation \& Results}\label{sec:experiments}

This section presents three experiments designed to explore the capabilities and limitations of the proposed methodology. Based on these results, possible applications are illustrated in Section \ref{sec:use-cases}, and future work paths are discussed in Section \ref{sec:discussion}. 

The first experiment in \S\ref{sec:iris-task} measures the capacity of the proposed model (IRIS-14B) at translating GIMPLE-to-LLVM IR on the evaluation sets described in \S\ref{sec:eval} (i.e., \textit{ExeBench-IRIS} and \textit{CodeForces-IRIS}). For context, IRIS-14B is benchmarked together with a wider variety of state-of-the-art open models. The second experiment, in \S\ref{sec:error-analysis}, explores the relationship between AI model failures and static metrics of code snippets. This allows the characterization of the code types that most frequently succeed or fail in the GIMPLE-to-LLVM IR translation task. Finally, \S\ref{sec:code-types} studies which data sources are most effective for training better LLM-based IR-to-IR translators.

\subsection{IRIS Task Evaluation} \label{sec:iris-task}
The first experiment measures models' capacity to solve the IRIS task, as defined in \S\ref{sec:iris-task}. Given a GIMPLE representation of an executable code, the goal is to produce the corresponding LLVM IR so that the code successfully compiles with LLVM and passes all associated tests. Benchmarked models include (i) IRIS-14B, the model produced in this work, and the only one explicitly trained for the task. IRIS-14B is also among the smallest model benchmarked. (ii) Qwen3-14B, the model used as a starting point for training IRIS-14B. (iii) Qwen3-Coder-A35B, a coder version of Qwen3, larger and more capable than the smaller 14B version. (iv) gpt-oss-120b, an OpenAI general-purpose open model, ranking top in most public leaderboards, (v) gpt-oss-20b, a smaller version of the same model for lower latency, and (vi) Kimi-K2-Instruct-0905, a general-purpose advanced model with one thousand billion parameters. Finally, we include (vii) LLM Compiler, a domain-specific model trained on a corpus of LLVM IR and assembly code and designed for compiler optimization tasks within the LLVM ecosystem.

Results for these experiments are reported in Table~\ref{tab:model_performance}. In general, \textit{Exebench-IRIS} is significantly easier for the models than \textit{CodeForces-IRIS}. This is most likely related to the nature of both benchmarks. As introduced in \S\ref{sec:eval}, \textit{ExeBench} samples, although extracted from real-world, repository-level code, have been simplified by the authors during the dataset processing. In contrast, \textit{CodeForces} contains a higher-quality feature selection due to its educational purpose and is generally more complex. On average, the \textit{Exebench-IRIS} sample has 20 lines of C code, while the average \textit{CodeForces} sample has 57 lines of C code. 

\begin{figure}[t]
    \centering

    \begin{subfigure}{\columnwidth}
        \centering
        \includegraphics[width=0.9\linewidth]{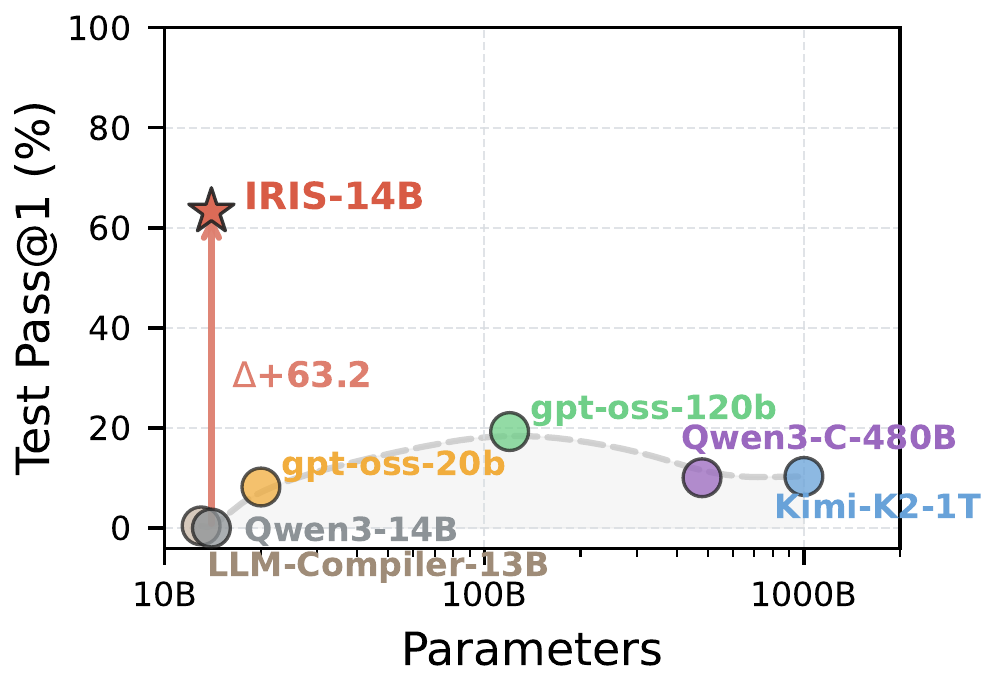}
        \caption{CodeForces-IRIS.}
        \label{fig:leaderboard-iris}
    \end{subfigure}

    \vspace{0.75em}

    \begin{subfigure}{\columnwidth}
        \centering
        \includegraphics[width=0.9\linewidth]{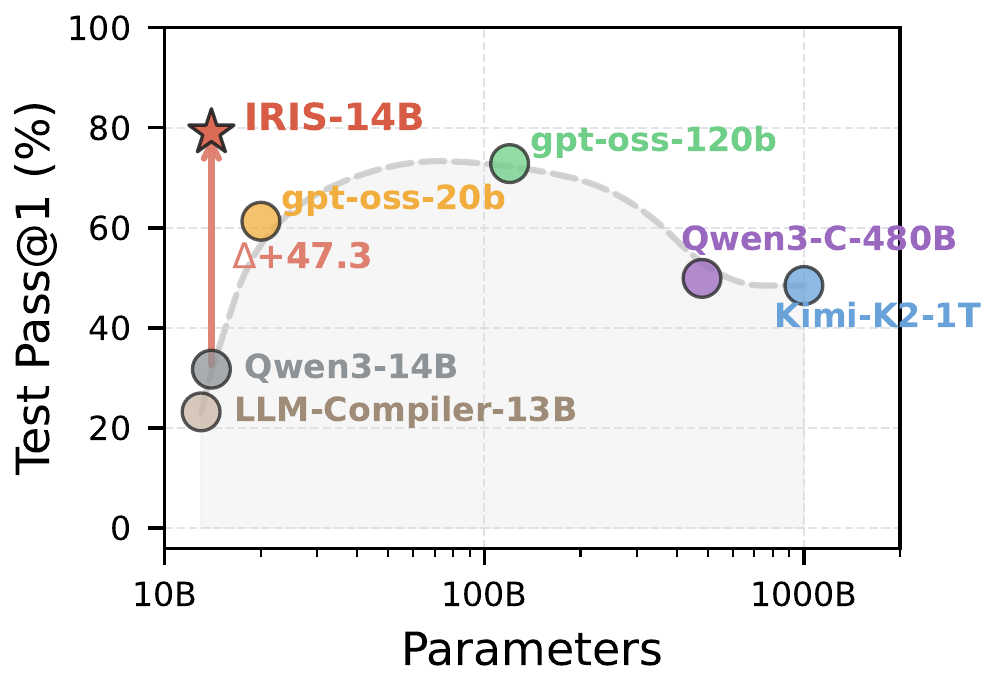}
        \caption{ExeBench-IRIS.}
        \label{fig:leaderboard-exebench}
    \end{subfigure}

    \caption{I/O test pass rate on the GIMPLE-to-LLVM IR translation task (vertical axis) vs model parameter size (horizontal axis, log-scaled) for six different models (colored circles) and IRIS-14B (star) on CodeForces-IRIS(a) and ExeBench-IRIS(b).}
    \label{fig:leaderboard}
\end{figure}

That being said, the IRIS task appears to be challenging even for large-scale state-of-the-art open models, which achieve only 50\% I/O Test pass rates on the easier \textit{Exebench-IRIS}. In fact, model size correlates weakly with task performance. This is represented in Figure \ref{fig:leaderboard}, which show the I/O pass rate with respect to the model size in billion parameters for \textit{CodeForces-IRIS} (in Figure \ref{fig:leaderboard-iris}) and \textit{Exebench-IRIS} (in Figure \ref{fig:leaderboard-exebench}). Despite having only 14B parameters, IRIS-14B outperforms widely used models with significantly larger parameter counts. While it's infeasible to quantify the number of IRs included in the crawled datasets used to train the benchmarked models, this quantity seems insufficient for both general-purpose and coder LLMs to solve the IRIS task. Interestingly, LLM Compiler surpasses IRIS-14B in compilation rate by +7\% on \textit{CodeForces-IRIS}, while its I/O test pass rate drops to below 1\%. This unusual pattern suggests that, although LLM Compiler is capable of generating LLVM IR that compiles, it fails to preserve the functional equivalence of the translated programs. This behavior is consistent with the model having been trained on a corpus of LLVM IR, which may allow it to produce syntactically valid IR without necessarily preserving the semantics of the source program. This observation also highlights the limitations of compilation-only metrics: a model could generate a trivial program that compiles successfully yet fails all functional tests. Therefore, the higher compilation rate of LLM Compiler does not indicate better IR translation quality.

Regarding the proposed model IRIS-14B, note that Qwen3-14B, the model it is based on, completely fails at the task, achieving the lowest success rates on both datasets across all models listed. In contrast, the training conducted to generate IRIS-14B consistently outperforms all baseline models on both benchmarks. On \textit{CodeForces-IRIS}, the proposed model achieves a compile success rate of 73.24\% and of 63.26\% in functional equivalence (I/O tests), improving almost 50 and 44 percentage points, respectively, over the strongest baseline (\textit{gpt-oss-120b}). On \textit{Exebench-IRIS}, IRIS-14B reaches 86.89\% compile success rate and 79.06\% I/O correctness, surpassing the best competing model by more than 6 percentage points on each metric. Overall, results show that task-specific fine-tuning on paired IR data is crucial for tackling IR-to-IR translation tasks, and that a specialized 14B-parameter model can outperform general models with up to two orders of magnitude more parameters.

\subsection{Error Analysis} \label{sec:error-analysis}

To better understand IRIS-14B's capabilities and limitations, this section examines the properties of the original C programs being translated that better characterize whether the model's translations will succeed or fail. This experiment focuses on the \textit{CodeForces-IRIS} test set because it proved more challenging for IRIS-14B (see Section \ref{sec:iris-task}). A more balanced behavior allows a clearer study of the relation between code features and translation success. For each submission in the test set, samples are labeled as \emph{successful} if the generated LLVM IR passes the syntactic correctness check, or \emph{failed} otherwise. The properties of the C programs considered are the 13 static metrics introduced in Section~\ref{sec:eval}.
  
\begin{figure*}
    \centering
\includegraphics[width=0.9\linewidth]{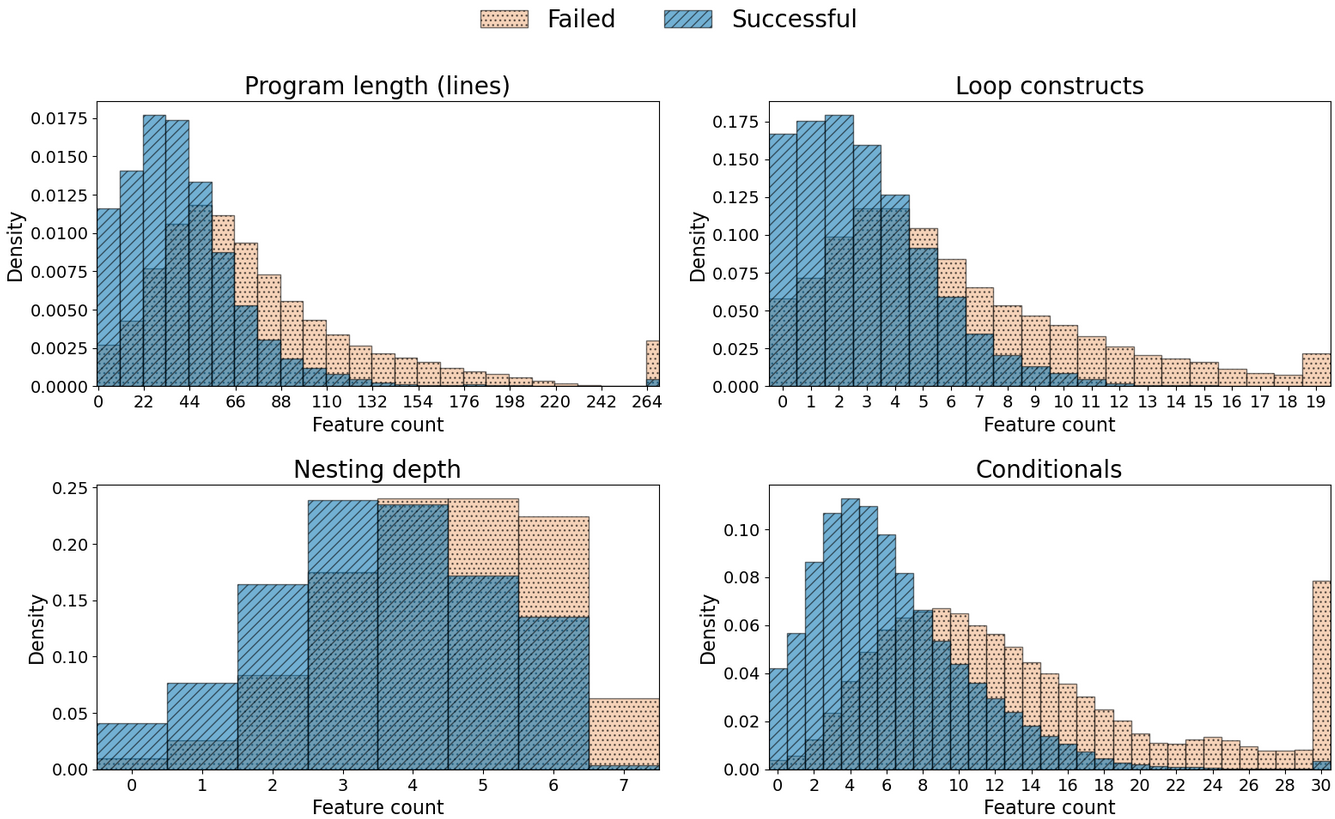}
    \caption{Distribution of four representative static metrics on the \textit{CodeForces} test set. Successful translations are shown in blue and failed translations in orange. The y-axis represents the normalized probability density.}
    \label{fig:static-metrics}
\end{figure*}

The first analysis investigates the distribution of static code metrics across the successful and failed test sample populations. Four of the most prevalent features in the dataset (present in more than 95\% of the samples) are plotted as histograms in Figure~\ref{fig:static-metrics}. These include metrics related to program size (lines) and control-flow complexity (loops, nesting depth, and conditionals). As shown in these plots, task success is strongly related to code complexity. For programs with less than 50 lines of code, less than 4 loop constructs, less than 3 nesting depth, and less than 8 conditionals, IRIS-14B has a success probability over 50\%. On the complementary cases (+50 lines, +5 loop constructs, +4 nesting depth, and +9 conditionals), IRIS-14B has a fail probability over 50\%. This marks the current state-of-the-art limit.

\begin{figure*}
    \centering
\includegraphics[width=0.99\linewidth]{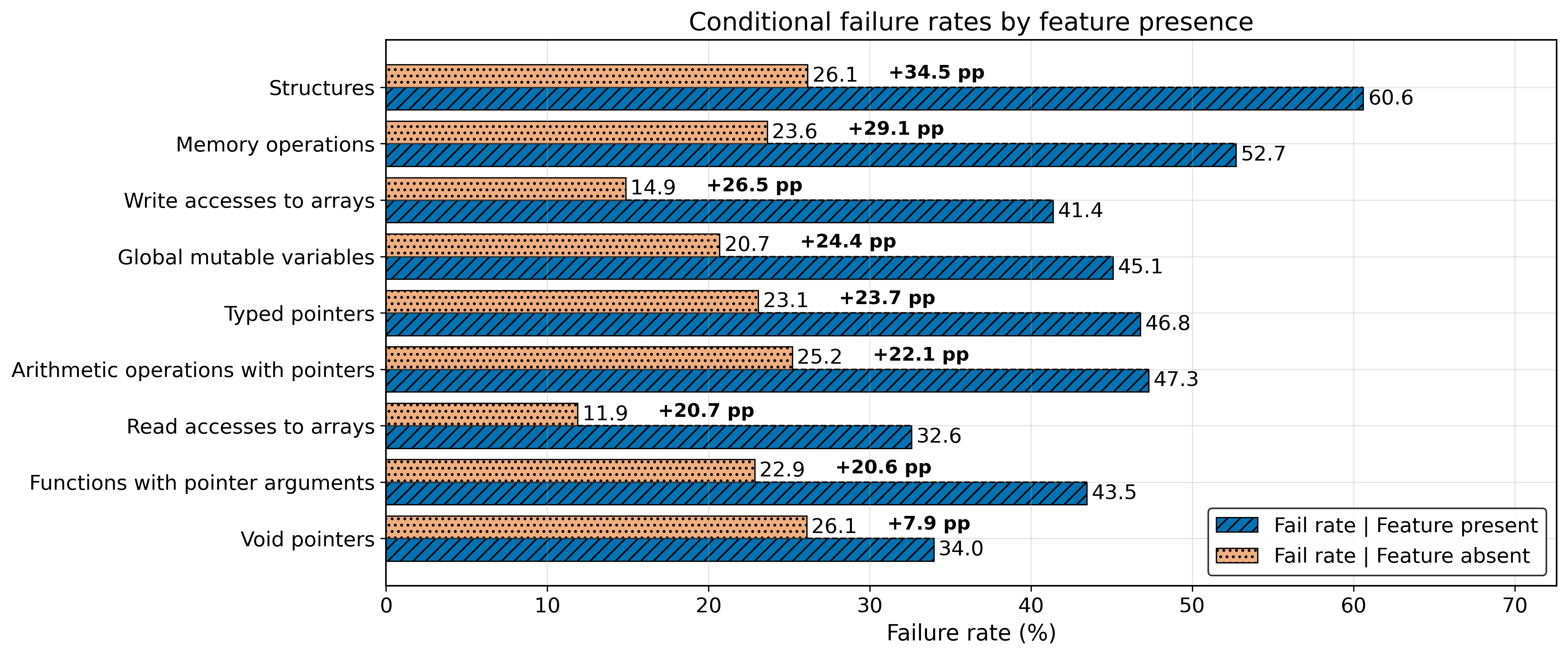}
    \caption{Conditional failure rates in percentages. Blue bars show the failure rate given that the feature is present in the dataset $P(\mathrm{fail}\mid f=1)$, whereas orange bars indicate the failure rate given that the feature is absent in the dataset $P(\mathrm{fail}\mid f=0)$. Sorted by the gap between both bars ($\Delta$, shown in bold and expressed in percentage points, pp), which indicates the impact of including a given code feature on the global failure rate on translation tasks.}
    \label{fig:conditional-rates}
\end{figure*}

To better characterize the relationship between the code features and task performance, a second analysis on feature presence is conducted. For each feature $f$, conditional failure rates are computed as $P(\mathrm{fail}\mid f) = \frac{F_f}{N_f}$, where $N_f$ accounts for the number of samples in which the feature is present, or absent, and $F_f$ the corresponding number of failed translations. $P(\mathrm{fail}\mid f=1)$ is computed as the failure rate (percentage) among samples in which the feature $f$ is present, and $P(\mathrm{fail}\mid f=0)$ as the failure rate among samples in which it is absent. Rather than comparing entire distributions, this analysis isolates the effect of individual features by measuring how the model’s failure rate changes when a given feature is present or absent in the input program. Figure \ref{fig:conditional-rates} presents this conditional perspective, allowing a more direct association of specific code features with increased or decreased translation difficulty.

Results in Figure \ref{fig:conditional-rates} are sorted by the difference between conditional fail rates to show a ranking of the most relevant features. This ranking is clearly led by the use of the \texttt{struct} keyword. The conditional presence of this feature increases the failure rate by almost +35\% points, well above that of any other feature.

The next group of challenging features induces a range of +25-30\% failure rates, and includes the use of memory operations involving functions such as \texttt{malloc()}, \texttt{free()}, and \texttt{memset()}, and write accesses to arrays, with the former being slightly harder than the latter (2.5\% more).

There is a third group of features associated with an increase in error rates of roughly +20-25\%. It includes pointer-related constructs (instantiation of typed, pointer arithmetic, and functions with pointer arguments), read accesses to arrays, as well as the use of mutable global variables. Finally, the presence of void pointers is  associated with an increase in error rates of roughly +8\%.

The major difference in error rates regarding structs is expected since they expose a fundamental difference between the IRs: while GIMPLE retains a high-level view of structs in which offsets and alignment are implicitly known via types, LLVM IR fully materializes them, requiring \texttt{getelementptr} to compute offsets and \texttt{load} operations to specify alignment. Our analysis further indicates that failures involving structs are most strongly associated with arrays of structs and struct-pointer usage. This analysis is consistent with the gradient of feature-error relationships of Figure \ref{fig:conditional-rates}, since array and pointer presence also accounts for model failures. Arrays also expose the implicit vs. explicit IR-flavors, but these are homogeneous and therefore easier to generalize by the model. Similarly, pointer operations are typically normalized into index-based accesses, behaving as arrays.

Overall, these findings suggest that translation difficulty increases with cumulative program complexity, while some constructs, such as \texttt{structs}, array write accesses, and memory operations entail remarkable complexity.

\subsection{Impact of Training Data} \label{sec:code-types}

The last experiment in this section concerns the importance of the code sources used to train the models for IR-to-IR translation. For this purpose, two models derived from the same base model (Qwen3-14B) are trained with an equal amount of data (82k samples), but using two different sources. These sources include the \textit{TheStack-IRIS} training dataset, presented in \S\ref{sec:training} and composed by a wide variety of public real-world code repositories, filtered by license and quality. The other training source used is \textit{CodeForces-IRIS}, presented in \S\ref{sec:eval}, which consists exclusively of competitive programming submissions. While \textit{TheStack} sources are likely to cover a broader range of instructions and operations, \textit{CodeForces} sources may exhibit higher average code quality due to its competitive programming setting. In this experiment, since \textit{CodeForces-IRIS} is used only in this specific setting for training, evaluation is conducted exclusively on the \textit{ExeBench-IRIS} dataset.

To avoid the effect of dataset size, we train two IRIS variants using the same number of training pairs: (i) 82k samples randomly selected from \textit{TheStack-IRIS} and (ii) 82k samples randomly selected from \textit{CodeForces-IRIS}. Both variants share the same model architecture, tokenization, and training procedure. Both models are evaluated on \textit{ExeBench-IRIS}, which is derived from real-world C code mined from GitHub repositories with a post-processing to ensure that samples remain manageable in size resembling competitive programming scenarios as discussed in Section \ref{sec:training}. The post-processing step is inherited from the original \textit{ExeBench} dataset and is not modified in our work.

Table~\ref{tab:code_types} reports the results using \texttt{pass@1}, measuring syntactic correctness through successful compilation and semantic correctness through I/O-based testing.

\begin{table}[t]
\centering
\caption{Accuracy in IR-to-IR translation (pass@1, N=1) for two IRIS variants trained with an equal budget of 82k samples obtained from different domains. Best in bold.} \label{tab:code_types}
\begin{tabular}{l|c|c}
&  \multicolumn{2}{c}{\textbf{Exebench}} \\
\cmidrule(lr){2-3}
\textbf{IRIS variants} & \textbf{Compile} (\%) & \textbf{I/O Test} (\%) \\
\midrule
IRIS-14B-CodeForces-82k & 68.67 & 59.92 \\
IRIS-14B-TheStack-82k   & \textbf{84.88} & \textbf{76.45}\\
\end{tabular}
\end{table}

The results show that the model trained on \textit{TheStack-IRIS} significantly outperforms the variant trained on \textit{CodeForces-IRIS} on both metrics. Even under a fixed number of training samples, exposure to heterogeneous, real-world code yields more robust IR-to-IR translation. This is likely the result of including code with a broader range of library usage patterns, control-flow shapes, and low-level operations encountered in real toolchains, a path for future dataset generation efforts to follow.
\section{Use-Cases} \label{sec:use-cases}

This work introduces a methodology for IR translation, together with dedicated training and evaluation methodologies and a model trained explicitly for the task. Beyond demonstrating the feasibility of this approach, it is essential to assess whether the IRIS-14B translation model generalizes to realistic and previously unsupported compilation scenarios. To this end, this section evaluates IRIS-14B on two out-of-distribution use-cases that exercise source languages and potentially IR features absent from the training data, thereby testing the model's robustness and practical applicability beyond the C-only setting.

The remainder of this section first shows that IRIS-14B enables the compilation of Ada programs that rely on the $Scalar\_Storage\_$\allowbreak$Order$ pragma using the LLVM toolchain. This pragma is supported by GNAT, the GCC-based Ada compiler, but remains unsupported in GNAT-LLVM, the LLVM-based version. Therefore, the example illustrates how IRIS can bridge feature gaps between existing frontends and backends without requiring changes in either component. Second, we show that IRIS-14B allows LLVM to process code written in Modula-2, a legacy language with mature GCC support but no native LLVM frontend. This example highlights the potential of LLM-based IR-to-IR translation mechanisms for tasks that currently lack native LLVM support.

\subsection{Support Ada's $Scalar\_Storage\_Order$ in LLVM}

The Ada programming language~\cite{ada2022WG9} has been closely associated with GCC since the 1990s through GNAT, an Ada frontend integrated in GCC. It has evolved over the decades to provide a full-featured Ada frontend, leveraging the mature optimization and backend infrastructure of the GCC framework. Efforts to introduce Ada support in LLVM were only initiated in the late 2010s, when AdaCore released GNAT-LLVM to use LLVM as an experimental backend. This work was motivated primarily by the need to access LLVM-only targets such as WebAssembly and LLVM's analysis ecosystem. Although GNAT-LLVM has evolved to support a richer set of Ada features, several features remain unsupported, including GCC-specific features or support for Ada2022 parallel constructs.

This use case considers the Ada program shown in Figure~\ref{lst:ada}. This code employs the $Scalar\_Storage$ $\_Order$ attribute, a GCC-specific feature that allows developers to control the byte order (endianness) of scalar components within composite types such as arrays or records, overriding the target machine’s default endianness to ensure consistent data representation across platforms. In the example, the $Scalar\_Storage\_Order$ representation clause is used to force a big-endian (high-order-first) byte layout for scalar elements of the $Byte\_Swapped\_Int\_Array$ type, even on little-endian architectures such as x86. This feature can be successfully compiled with GNAT, but it fails when using the LLVM framework because GNAT-LLVM lacks support for it.

\begin{figure}[h]
\begin{lstlisting}[language=Ada, basicstyle=\small]
with System;

function main return Integer is
   type Byte_Swapped_Int_Array is array (1..1) of \
   Integer;
   for Byte_Swapped_Int_Array'Scalar_Storage_Order \
   use System.High_Order_First;

   X : Byte_Swapped_Int_Array := (1 => 30);
   Y : Integer;
begin
   Y := X(1);
   return Y;
end main;
\end{lstlisting}
\caption{Ada code using the Scalar\_Storage\_Order attribute.} \label{lst:ada}
\end{figure}

Following the GIMPLE-to-LLVM IR translation methodology, the GNAT frontend first parses and type-checks the source code, and then lowers the program to GIMPLE as part of its standard compilation pipeline. Rather than continuing the compilation to RTL and generating code for a GCC-supported architecture, we intercept the process at the GIMPLE level and provide the resulting GIMPLE dump as input to IRIS-14B. IRIS-14B then translates the GIMPLE source into semantically equivalent LLVM IR, which is passed to the LLVM toolchain to produce the final executable.

This pipeline enables Ada programs that rely on GCC-specific features to be compiled within the LLVM ecosystem, without modifying the existing GNAT-LLVM Ada frontend or changing the original source code. As such, this use case illustrates IRIS-14B as a practical interoperability mechanism, capable of extending the applicability of LLVM to existing codebases and language features without reimplementing frontends.

\subsection{Compile Modula-2 Programs with LLVM}
Modula-2 is a strongly typed systems programming language introduced in the late 1970s as a modular successor to Pascal~\cite{Wirth1982Modula2}. It provides explicit module constructs for separate compilation and information hiding, together with low-level features suitable for systems programming. Historically, Modula-2 has been used in operating systems, compiler research, and embedded and real-time systems. In contemporary toolchains, active support for Modula-2 is primarily provided through the GCC-based \texttt{gm2} frontend, while no native Modula-2 frontend exists in the LLVM ecosystem.

\begin{figure*}
    \centering
    \includegraphics[width=0.99\linewidth]{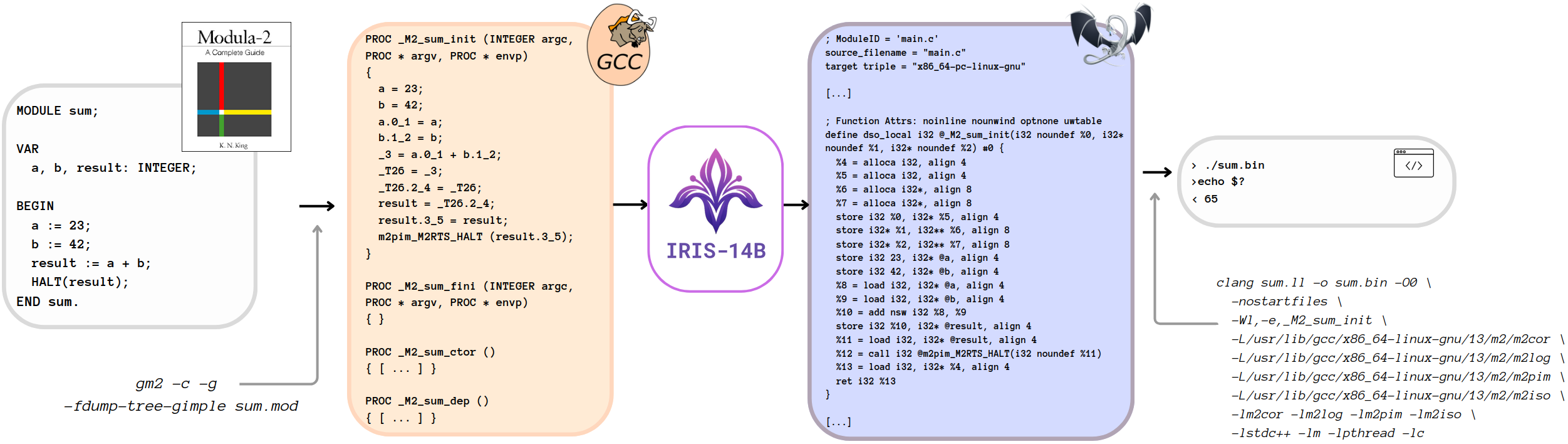}
    \caption{Workflow for the Modula-2 use case. A Modula-2 source program is provided as input to the pipeline. Using the existing GCC compiler toolchain, we first extract the corresponding GIMPLE IR, which IRIS-14B then translates to LLVM IR. The resulting LLVM IR enables compilation of the original program with the LLVM toolchain, linking against the Modula-2 runtime, a capability not supported by existing compilation tools.}
    \label{fig:modula2-use-case}
\end{figure*}

This setting constitutes a representative use case for IRIS-14B. By translating the GIMPLE emitted by the GCC Modula-2 frontend into semantically equivalent LLVM IR, IRIS-14B enables Modula-2 programs to be processed by the LLVM toolchain. This, in turn, allows legacy Modula-2 codebases to benefit from modern LLVM-based tooling and backends, such as additional compilation targets, sanitizers, and optimization passes, without requiring a dedicated Modula-2 frontend for LLVM. Figure~\ref{fig:modula2-use-case} presents the function-level proof of concept used for this experiment, a Modula-2 code snippet computing simple arithmetics. This code has been successfully compiled using the proposed methodology for GCC and LLVM integration, employing GIMPLE-to-LLVM IR translation with IRIS-14B.

\section{Discussion}\label{sec:discussion}

This work shows how transformer models are suitable for GIMPLE-to-LLVM IR translation, eliminating the need for hand-crafted rule-based methods that require costly maintenance. The goal is achieved through the proposed methodology, which involves supervised fine-tuning on curated datasets. The results analyzed in Section~\ref{sec:experiments} raise several important considerations, which outline directions for future work and are worthy of the discussions presented next.

\subsection{Language-dependent IR features}\label{sec:language-dependency}

IRIS-14B is trained on paired GIMPLE and LLVM IR samples obtained by compiling the same C source code with GCC and Clang. We acknowledge that other programming languages, such as C++ or Fortran, which are mature and consistently supported across both toolchains, are also suitable candidates for model training under our methodology. However, restricting the dataset to C snippets provides a controlled and reliable training set, making it easier to attribute successes and failures primarily to the IR-to-IR translation problem rather than to ambiguities introduced by cross-language semantic mismatches.

There is no notion of language-dependent constructs in the LLVM IR documentation~\cite{llvm_langref}. Even in C++, where higher-level constructs such as classes, inheritance, and exception-handling semantics exist and are absent from C, these are ultimately lowered to LLVM IR constructs that are already present in the IR derived from C codes. For instance, the LLVM IR type system includes source-language-independent primitive types and only four derived types: pointers, arrays, structures, and functions~\cite{lattner2004llvm}. Consequently, C++ classes with inheritance are lowered to lower-level entities such as structure types for object layout and functions for methods. We also observe the lowering of high-level C++ features into common IR constructs in GIMPLE. However, we find less strict policies in GIMPLE. For example, the type system appears less canonicalized, as illustrated by the handling of boolean types: values originating from C are represented as \texttt{\_Bool}, whereas those from C++ are represented as \texttt{bool}.

Although LLVM IR and GIMPLE rely on sets of language-independent constructs, the generated IR may still expose frontend lowering and ABI-specific patterns. Indeed, C++ code can introduce mangled symbol names, anonymous namespaces, and runtime-library calls associated with the target ABI. These are not C++-specific IR constructs, but language-specific encodings built on the same underlying IR representation. For example, GIMPLE generated from empty C++ functions typically includes a \texttt{GIMPLE\_NOP} statement, whereas empty C functions lowered to GIMPLE generally retain an empty function body.

Despite these differences in high-level source-language features and source-dependent lowering patterns, the common IR structural representation remains, and it is captured by IRIS-14B. As an illustrative example, Figure~\ref{fig:overload-cpp} shows a C++ snippet implementing method overloading, a feature specific to C++. In this case, the model receives as input the GIMPLE representation shown in Figure~\ref{fig:overload-gimple}, where the two overloads appear as separate function forms with different parameter lists, and both targets are directly called from \texttt{main}. The ground-truth LLVM IR shown in Figure~\ref{fig:overload-llvm-gt} likewise represents the two overloads as distinct functions with different mangled names and signatures, and \texttt{main} calls each of them explicitly. Although the model was trained on GIMPLE lowered from C, where method overloading does not exist, the generated LLVM IR (Figure~\ref{fig:overload-llvm-gen}) still captures the core overloading pattern by emitting two distinct functions with different parameter lists and by calling each method explicitly from \texttt{main}. However, unlike the ground-truth LLVM IR, it does not encode the overloads using ABI-level C++ name mangling. Instead, as in GIMPLE, it preserves source-like tokens such as \texttt{@A::f} for both functions, causing a symbol collision in the LLVM module. The issue can be resolved by assigning each overload a unique IR-level symbol name.

\begin{figure*}[t]
\centering

\begin{subfigure}[t]{0.48\textwidth}
\centering
\begin{lstlisting}[language=C++, basicstyle=\ttfamily\small]
class A {
public:
    void f() {}
    void f(int) {}
};

int main() {
    A a;
    a.f();
    a.f(1);
}
\end{lstlisting}
\caption{C++ source code illustrating method overloading. The two definitions of \texttt{f} differ only in their parameter lists, and overload resolution is expressed in the source directly through the call parameters.}
\label{fig:overload-cpp}
\end{subfigure}\hfill
\begin{subfigure}[t]{0.48\textwidth}
\centering
\begin{lstlisting}[language=C++, basicstyle=\ttfamily\small]
int main () {
  struct A a;
  A::f (&a);
  A::f (&a, 1);
}

void A::f (struct A *this)
{ GIMPLE_NOP }

void A::f (struct A *this, int)
{ GIMPLE_NOP }
\end{lstlisting}
\caption{GIMPLE representation as emitted by GCC. The source-level overload has been resolved into separate functions with different parameter lists. Functions are encoded using the same qualified name \texttt{A::f}, reflecting their association with the class, while the implicit receiver is made explicit through \texttt{this}.}\label{fig:overload-gimple}
\end{subfigure}

\vspace{0.5em}

\begin{subfigure}[t]{0.48\textwidth}
\centering
\begin{lstlisting}[basicstyle=\ttfamily\small]
define i32 @main() {
  call void @_ZN1A1fEv(ptr %a)
  call void @_ZN1A1fEi(ptr %a, i32 1)
  ret i32 0
}

define void @_ZN1A1fEv(ptr %0) { ... }
define void @_ZN1A1fEi(ptr %0, i32 %1) { ... }
\end{lstlisting}
\caption{Ground-truth LLVM IR extracted from the LLVM toolchain. Method overloading is lowered to two distinct functions distinguished by both their signatures and their ABI-mangled C++ symbol names, \texttt{\_ZN1A1fEv} and \texttt{\_ZN1A1fEi}. The calls in \texttt{main} explicitly target each resolved overload.}\label{fig:overload-llvm-gt}
\end{subfigure}\hfill
\begin{subfigure}[t]{0.48\textwidth}
\centering
\begin{lstlisting}[basicstyle=\ttfamily\small]
define i32 @main() {
  call void @A::f(ptr %a)
  call void @A::f(ptr %a, i32 1)
  ret i32 0
}

define void @A::f(ptr %0) { ... }
define void @A::f(ptr %0, i32 %x) { ... }
\end{lstlisting}
\caption{Model-generated LLVM IR, which preserves the core overloading pattern by emitting two distinct functions and explicit calls, but uses source-like (identical) names instead of C++ ABI mangled symbols.}\label{fig:overload-llvm-gen}
\end{subfigure}

\caption{Method overloading across representations. Although method overloading is a C++-specific feature absent from the C-derived training data, IRIS-14B reproduces its core structural lowering pattern from GIMPLE to LLVM IR. In particular, the model correctly emits separate functions for each overload and explicit calls to the resolved targets, while differing from the ground truth in its use of source-like symbol names.}\label{fig:cpp-overloading}
\end{figure*}

This model behavior is due to the C-only training. To incorporate language-specific artifacts such as mangled names, we are exploring continual-learning approaches that would enable the model to adapt to language-specific lowering without retraining it from scratch. In particular, we envision lightweight adaptation strategies that can enable the model to recognize source-specific patterns while preserving its existing capabilities, thereby broadening language coverage and improving generalization. Extending this coverage will also require large-scale evaluation on such out-of-distribution languages.

Similarly, the IR constructs that the model learned from the C-based IR produced in this work may still not cover the full range of IR semantics in LLVM IR and GIMPLE. For example, certain corner-case implementations in C may be absent from the training data, and some IR patterns may be triggered more frequently by specific source-language constructs than by others. Extending the model with the post-training strategies mentioned would also help capture these underrepresented constructs and enhance the generalization and accuracy of the model.

\subsection{Compiler's evolution}

The evolution of the compiler toolchains over time may introduce changes in the IRs used in this work. However, such changes are typically governed by design policies that prioritize practical compatibility across versions. LLVM’s developer policy indicates that while the textual IR format itself is not guaranteed to be strictly backward compatible, the toolchain aims to preserve practical compatibility when evolving the IR~\footnote{\url{https://llvm.org/docs/DeveloperPolicy.html\#ir-backwards-compatibility}.\newline Last accessed May 2026.}. In particular, newer LLVM releases are expected to load older bitcode versions and upgrade them when necessary, ensuring that legacy constructs are not miscompiled even if some features are deprecated or dropped during the upgrade process.

Consistent with this policy, most IR changes in recent LLVM releases primarily introduce incremental refinements, rather than large redesigns. For example, LLVM 18 removed several legacy constant-expression forms (e.g., \texttt{and}, \texttt{or}, \texttt{zext}, \texttt{fptosi})~\footnote{\url{https://releases.llvm.org/18.1.0/docs/ReleaseNotes.html\#changes-to-the-llvm-ir}.\newline Last accessed May 2026.}, LLVM 19 continued this cleanup by removing constant-expression variants of \texttt{icmp}, \texttt{fcmp}, and \texttt{shl}~\footnote{\url{https://releases.llvm.org/19.1.0/docs/ReleaseNotes.html\#changes-to-the-llvm-ir}.\newline Last accessed May 2026.}, and LLVM 21 further removed the constant-expression form of \texttt{mul}~\footnote{\url{https://releases.llvm.org/21.1.0/docs/ReleaseNotes.html\#changes-to-the-llvm-ir}.\newline Last accessed May 2026.}. Other changes are similarly localized, typically affecting specific instructions, intrinsics, or attributes without altering the underlying memory or control-flow semantics: LLVM 21 replaced the \texttt{nocapture} attribute with \texttt{captures(none)}, LLVM 19 renamed several vector intrinsics from the \texttt{llvm.experimental.*} namespace to their stable \texttt{llvm.*} forms, and LLVM 22 changed the interface of masked memory intrinsics by moving alignment information from an explicit operand to a pointer attribute~\footnote{\url{https://releases.llvm.org/22.1.0/docs/ReleaseNotes.html\#changes-to-the-llvm-ir}.\newline Last accessed May 2026.}. Even major transitions, such as the shift from typed to opaque pointers finalized in LLVM 17, preserve enough structural continuity for older IR to remain processable by newer toolchains. The GIMPLE representation, although less thoroughly documented, also appears to be even more stable across GCC versions. Notably, we compared the GIMPLE sections of the GCC internals manuals for GCC 11 (2023)~\footnote{\url{https://gcc.gnu.org/onlinedocs/gcc-11.4.0/gccint.pdf}. Last accessed May 2026.} and GCC 15 (2025)~\footnote{\url{https://gcc.gnu.org/onlinedocs/gcc-15.2.0/gccint.pdf}. Last accessed May 2026.} and found only one addition: \texttt{GIMPLE\_OMP\_STRUCTURED\_BLOCK}, introduced as a new tuple-specific accessor related to OpenMP.

Under this perspective, we empirically evaluate how such evolution impacts IR-to-IR translation. During the development of IRIS-14B, we trained an early version of the model using GIMPLE and LLVM IR textual dumps generated by GCC 11.4.0 and LLVM 14.0.0, respectively. This LLVM version predates the transition to opaque pointers, which became the canonical representation starting in LLVM 17~\footnote{\url{https://releases.llvm.org/17.0.1/docs/ReleaseNotes.html\#changes-to-the-llvm-ir}. Last accessed May 2026.}. Despite this difference in compiler versions, the model achieved performance comparable to the results reported in this paper, which uses newer toolchain versions (GCC 15 and LLVM 22). Furthermore, the IR generated by this earlier model can be successfully compiled and evaluated using not only LLVM 14, but also the latest LLVM 22. These results indicate that the structural patterns learned by the model from older IR versions remain valid across compiler versions, as even major transitions such as the introduction of opaque pointers do not prevent newer compilers from correctly compiling older IR representations. This behavior is consistent with the compiler's usual practice of maintaining backward compatibility. At the same time, although older constructs remain valid, the introduction of new IR features requires updating the model to recognize them, as is common in conventional compiler tooling.

Consequently, while IRs can evolve with time, the results of this work suggest that many of the core patterns present in GIMPLE and LLVM IR remain sufficiently stable across compiler versions to support data-driven approaches such as IRIS-14B and that deviations can be successfully incorporated into the model's knowledge through targeted post-training strategies as discussed in \S\ref{sec:language-dependency} for out-of-distribution languages.

From a broader perspective, the findings of this study point to a role for LLMs in future compiler stacks, not as replacements for traditional compilers, but as complementary components within hybrid neuro-symbolic systems. In particular, IRIS-14B can be understood as an interoperability layer for cross-toolchain workflows, enabling interaction between compiler ecosystems without requiring modifications to existing compiler passes. Under this design, the fast and deterministic compiler infrastructure remains responsible for conventional compilation and optimization, while the AI model is used only for steps that current toolchains cannot readily support, such as cross-ecosystem IR translation or optimization pass-ordering prediction for application-specific workloads in which data-driven alternatives overcome the weaknesses of rule-based methods. This vision aligns with recent work~\cite{zhang2026new} that similarly highlights hybrid compiler architectures as one of the most promising near-term directions for the convergence of LLMs and compilers.

\subsection{LLVM IR-to-GIMPLE translation}\label{sec:direction}

This work focuses on the GIMPLE-to-LLVM IR translation direction, as only the LLVM toolchain reliably supports starting compilation from an IR dump. From a practical perspective, however, the reverse direction, from LLVM IR to GIMPLE, may also be widely used. Many modern languages, such as Rust, are natively developed for LLVM, which offers a modular, widely adopted infrastructure. In contrast, GCC continues to provide strong support for certain embedded and legacy architectures that LLVM does not cover. Enabling translation from LLVM IR to GIMPLE would therefore open a path for LLVM-based frontends to target GCC-only platforms. In addition, the active development of a GCC-based Rust compiler, e.g., \texttt{gccrs}~\cite{gccrs}, also underscores that there is concrete demand for targeting GCC-only platforms from currently LLVM-based languages like Rust, and this support would also enable richer compiler testing and verification via cross-checking behavior across GCC and LLVM.

As a first step in this direction, we investigated how to use the experimental \textit{GIMPLE FrontEnd}~\cite{gcc_gimple_fe} to start compilation from GIMPLE dumps. GIMPLE FrontEnd currently accepts only a subset of GIMPLE, primarily for unit testing and debugging purposes. To better align GCC’s internal dumps with this parser, GCC provides a GIMPLE dump modifier \texttt{-gimple}. When combined with standard dump flags, it produces tree dumps that more closely match the format accepted by the GIMPLE parser. For example, \texttt{-fdump-tree-gimple} becomes \texttt{-fdump-tree-gimple-gimple}.

The GIMPLE FrontEnd is enabled via the \texttt{-fgimple} option. It uses the $\texttt{\_\_GIMPLE}$ annotation on functions to indicate that their bodies are written directly in GIMPLE rather than in C. The $\texttt{\_\_GIMPLE}$ parser is integrated with the C tokenizer and preprocessor, and the optional \texttt{startwith} argument allows the user to specify the compiler pass at which processing should begin.

In practice, some post-processing of these dumps is still required before they can be successfully re-ingested by GIMPLE FrontEnd. Developing a robust LLVM-to-GIMPLE IR translation pipeline that interoperates with this mechanism is part of our ongoing work.

\subsection{Context Length Limitations} \label{sec:context}
One of the main limitations of current LLMs is their restricted context window, i.e., the maximum input sequence length that a model can concurrently consider. In this work, we use a reasonably long 32k-token context (1 token corresponds to approximately 4 characters), which is sufficient to include real-world code samples from \textit{Exebench-IRIS} as well as competitive programming problems from \textit{CodeForces-IRIS}, where programs are typically self-contained. However, for raw repositories such as the \textit{GNU utils}, where dependencies span multiple files and source files can be thousands of lines long, full-program translation often exceeds the available context.

Differences in IR verbosity further influence this challenge. In practice, LLVM IR representations are substantially more verbose than their GIMPLE counterparts, with LLVM IR requiring, on average, approximately three times as many tokens as GIMPLE for an equivalent source code functionality. As a result, even when GIMPLE inputs fit comfortably within the model’s context window, the corresponding LLVM IR outputs (unknown during inference time, as the response is being decoded token by token) may approach or exceed context limits.

During training, we mitigated this issue by parsing GIMPLE and LLVM IR dumps and extracting function-level pairs, as illustrated in Figure~\ref{fig:iris-gnu}. However, these samples do not compile as standalone units and therefore do not form complete translation units. In practice, this does not appear to have a substantial impact on model performance, since the majority of training samples (approximately 96\%) correspond to complete translation units, while the remaining function-level samples provide additional training diversity due to the nature of the source corpora from which they are derived. Context length limit is an explicit limitation of the current technology, as applications of high interest are typically larger programs and may yet exhibit different translation challenges, such as an increasing number of feature counts, which we found associated with failure rates in \S\ref{sec:error-analysis}.

For further addressing this issue, one option is to consider recent methods for extending the model's context length~\cite{su2024roformer}. Another path would be to decompose long, dependency-rich translation problems into smaller tasks that the model can process within its context limits.
\section{Conclusions}

This work takes the first steps toward learning-based compiler interoperability by introducing IRIS-14B, the first transformer model specially trained for GIMPLE-to-LLVM IR translation. While the scale of the training dataset plays an important role, the selection of the data is also found to be fundamental to achieving better model success rates. In addition to the training sets, this work also releases the two evaluation sets produced. 

The experiments performed on the proposed open-source model, trained to translate GIMPLE to LLVM IR, suggest that data-driven methods might overcome the limitations that rule-based approaches have faced over the decades. Across competitive programming and real-world code, IRIS-14B achieves high syntactic and functional correctness, consistently outperforming the larger state-of-the-art general-purpose and coding models. IRIS-14B demonstrates that LLMs can learn rich, semantics-preserving mappings between heterogeneous compiler IRs.

This work enables a practical methodology for integrating LLM-based IR translation into existing compiler toolchains without modifying existing frontends or backends. This means new compilation workflows, including the reuse of LLVM backends and tooling for languages primarily supported by GCC. The applicability of this approach is demonstrated by compiling GCC-only Ada features and GCC-only languages such as Modula-2 with the LLVM toolchain.

Overall, our results demonstrate that IR-to-IR translation is a viable application of LLMs. One that enables more modular and interoperable compiler infrastructure by combining AI-based components with traditional compiler toolchains. The paths for future work in this direction are many, and of high interest to many.


\section*{Acknowledgments}
This work was partially funded by the HiPART project, with reference \texttt{PID2023-148117NA-I00}, financed by \texttt{MICIU/AEI/10.13039/}\allowbreak\texttt{501100011033} and FEDER, UE. Additionally, this work was partially supported by the ELLIOT project funded by the European Union under grant agreement No. \texttt{101214398}, and by project \texttt{PID2023}\allowbreak\texttt{-146511NBI00} funded by the Spanish Ministry of Science, Innovation and Universities \texttt{MCIU/AEI/10.13039/}\allowbreak\texttt{501100011033}, and by the EU ERDF. Finally, this work was also supported by the AI4S fellowships awarded to Andrea Valenzuela and Cristian Gutierrez fellowships within the “Generación D” initiative, Red.es, Ministerio para la Transformación Digital y de la Función Pública, for talent attraction (\texttt{C005/24-ED CV1}). Funded by the European Union NextGenerationEU funds, through PRTR. We also thank Adrian Munera for his valuable insights into the LLVM toolchain.

\bibliographystyle{ACM-Reference-Format}
\bibliography{text/refs}


\begin{thebibliography}{45}


\ifx \showCODEN    \undefined \def \showCODEN     #1{\unskip}     \fi
\ifx \showISBNx    \undefined \def \showISBNx     #1{\unskip}     \fi
\ifx \showISBNxiii \undefined \def \showISBNxiii  #1{\unskip}     \fi
\ifx \showISSN     \undefined \def \showISSN      #1{\unskip}     \fi
\ifx \showLCCN     \undefined \def \showLCCN      #1{\unskip}     \fi
\ifx \shownote     \undefined \def \shownote      #1{#1}          \fi
\ifx \showarticletitle \undefined \def \showarticletitle #1{#1}   \fi
\ifx \showURL      \undefined \def \showURL       {\relax}        \fi
\providecommand\bibfield[2]{#2}
\providecommand\bibinfo[2]{#2}
\providecommand\natexlab[1]{#1}
\providecommand\showeprint[2][]{arXiv:#2}

\bibitem[9(2022)]%
        {ada2022WG9}
\bibfield{author}{\bibinfo{person}{Ada Working Group ISO/IEC JTC 1/SC~22/WG
  9}.} \bibinfo{year}{2022}\natexlab{}.
\newblock \bibinfo{title}{{Ada Reference Manual}}.
\newblock
\newblock
\shownote{\url{http://www.ada-auth.org/standards/22rm/RM-Final.pdf}}.


\bibitem[Achiam et~al\mbox{.}(2023)]%
        {achiam2023gpt}
\bibfield{author}{\bibinfo{person}{Josh Achiam}, \bibinfo{person}{Steven
  Adler}, \bibinfo{person}{Sandhini Agarwal}, \bibinfo{person}{Lama Ahmad},
  \bibinfo{person}{Ilge Akkaya}, \bibinfo{person}{Florencia~Leoni Aleman},
  \bibinfo{person}{Diogo Almeida}, \bibinfo{person}{Janko Altenschmidt},
  \bibinfo{person}{Sam Altman}, \bibinfo{person}{Shyamal Anadkat},
  {et~al\mbox{.}}} \bibinfo{year}{2023}\natexlab{}.
\newblock \showarticletitle{{GPT-4 technical report}}.
\newblock \bibinfo{journal}{\emph{arXiv preprint arXiv:2303.08774}}
  (\bibinfo{year}{2023}).
\newblock


\bibitem[Agarwal et~al\mbox{.}(2025)]%
        {agarwal2025gpt}
\bibfield{author}{\bibinfo{person}{Sandhini Agarwal}, \bibinfo{person}{Lama
  Ahmad}, \bibinfo{person}{Jason Ai}, \bibinfo{person}{Sam Altman},
  \bibinfo{person}{Andy Applebaum}, \bibinfo{person}{Edwin Arbus},
  \bibinfo{person}{Rahul~K Arora}, \bibinfo{person}{Yu Bai},
  \bibinfo{person}{Bowen Baker}, \bibinfo{person}{Haiming Bao},
  {et~al\mbox{.}}} \bibinfo{year}{2025}\natexlab{}.
\newblock \showarticletitle{gpt-oss-120b \& gpt-oss-20b model card}.
\newblock \bibinfo{journal}{\emph{arXiv preprint arXiv:2508.10925}}
  (\bibinfo{year}{2025}).
\newblock


\bibitem[Armengol-Estap{\'e} et~al\mbox{.}(2022)]%
        {armengol2022exebench}
\bibfield{author}{\bibinfo{person}{Jordi Armengol-Estap{\'e}},
  \bibinfo{person}{Jackson Woodruff}, \bibinfo{person}{Alexander Brauckmann},
  \bibinfo{person}{Jos{\'e} Wesley de~Souza Magalhaes}, {and}
  \bibinfo{person}{Michael~FP O'Boyle}.} \bibinfo{year}{2022}\natexlab{}.
\newblock \showarticletitle{ExeBench: an ML-scale dataset of executable C
  functions}. In \bibinfo{booktitle}{\emph{Proceedings of the 6th ACM SIGPLAN
  International Symposium on Machine Programming}}. \bibinfo{pages}{50--59}.
\newblock


\bibitem[Bai et~al\mbox{.}(2023)]%
        {bai2023qwen}
\bibfield{author}{\bibinfo{person}{Jinze Bai}, \bibinfo{person}{Shuai Bai},
  \bibinfo{person}{Yunfei Chu}, \bibinfo{person}{Zeyu Cui},
  \bibinfo{person}{Kai Dang}, \bibinfo{person}{Xiaodong Deng},
  \bibinfo{person}{Yang Fan}, \bibinfo{person}{Wenbin Ge}, \bibinfo{person}{Yu
  Han}, \bibinfo{person}{Fei Huang}, {et~al\mbox{.}}}
  \bibinfo{year}{2023}\natexlab{}.
\newblock \showarticletitle{{Qwen technical report}}.
\newblock \bibinfo{journal}{\emph{arXiv preprint arXiv:2309.16609}}
  (\bibinfo{year}{2023}).
\newblock


\bibitem[Cummins et~al\mbox{.}(2023)]%
        {cummins2023large}
\bibfield{author}{\bibinfo{person}{Chris Cummins}, \bibinfo{person}{Volker
  Seeker}, \bibinfo{person}{Dejan Grubisic}, \bibinfo{person}{Mostafa
  Elhoushi}, \bibinfo{person}{Youwei Liang}, \bibinfo{person}{Baptiste
  Roziere}, \bibinfo{person}{Jonas Gehring}, \bibinfo{person}{Fabian Gloeckle},
  \bibinfo{person}{Kim Hazelwood}, \bibinfo{person}{Gabriel Synnaeve},
  {et~al\mbox{.}}} \bibinfo{year}{2023}\natexlab{}.
\newblock \showarticletitle{{Large language models for compiler optimization}}.
\newblock \bibinfo{journal}{\emph{arXiv preprint arXiv:2309.07062}}
  (\bibinfo{year}{2023}).
\newblock


\bibitem[Cummins et~al\mbox{.}(2025)]%
        {cummins2025llm}
\bibfield{author}{\bibinfo{person}{Chris Cummins}, \bibinfo{person}{Volker
  Seeker}, \bibinfo{person}{Dejan Grubisic}, \bibinfo{person}{Baptiste
  Roziere}, \bibinfo{person}{Jonas Gehring}, \bibinfo{person}{Gabriel
  Synnaeve}, {and} \bibinfo{person}{Hugh Leather}.}
  \bibinfo{year}{2025}\natexlab{}.
\newblock \showarticletitle{Llm compiler: Foundation language models for
  compiler optimization}. In \bibinfo{booktitle}{\emph{Proceedings of the 34th
  ACM SIGPLAN International Conference on Compiler Construction}}.
  \bibinfo{pages}{141--153}.
\newblock


\bibitem[Eniser et~al\mbox{.}(2024)]%
        {eniser2024towards}
\bibfield{author}{\bibinfo{person}{Hasan~Ferit Eniser},
  \bibinfo{person}{Hanliang Zhang}, \bibinfo{person}{Cristina David},
  \bibinfo{person}{Meng Wang}, \bibinfo{person}{Maria Christakis},
  \bibinfo{person}{Brandon Paulsen}, \bibinfo{person}{Joey Dodds}, {and}
  \bibinfo{person}{Daniel Kroening}.} \bibinfo{year}{2024}\natexlab{}.
\newblock \showarticletitle{{Towards translating real-world code with LLMs: A
  study of translating to Rust}}.
\newblock \bibinfo{journal}{\emph{arXiv preprint arXiv:2405.11514}}
  (\bibinfo{year}{2024}).
\newblock


\bibitem[GCC(2025)]%
        {gcc25dragonegg}
\bibfield{author}{\bibinfo{person}{GCC}.} \bibinfo{year}{2025}\natexlab{}.
\newblock \bibinfo{title}{{DragonEgg}}.
\newblock
\newblock
\shownote{\url{https://dragonegg.llvm.org/}}.


\bibitem[{GCC Developer Community}(2019)]%
        {gcc_gimple_fe}
\bibfield{author}{\bibinfo{person}{{GCC Developer Community}}.}
  \bibinfo{year}{2019}\natexlab{}.
\newblock \bibinfo{title}{GIMPLE FE: A Gimple Front End}.
\newblock
  \bibinfo{howpublished}{\url{https://gcc.gnu.org/wiki/GimpleFrontEnd}}.
\newblock
\newblock
\shownote{Accessed: 11 December 2025}.


\bibitem[GitHub(2025)]%
        {github25copilot}
\bibfield{author}{\bibinfo{person}{GitHub}.} \bibinfo{year}{2025}\natexlab{}.
\newblock \bibinfo{title}{{Copilot}}.
\newblock
\newblock
\shownote{\url{https://github.com/copilot}}.


\bibitem[{GNU Project}(2026)]%
        {gnu_packages}
\bibfield{author}{\bibinfo{person}{{GNU Project}}.}
  \bibinfo{year}{2026}\natexlab{}.
\newblock \bibinfo{title}{GNU Software}.
\newblock
\urldef\tempurl%
\url{https://www.gnu.org/software/software.html#allgnupkgs}
\showURL{%
\tempurl}
\newblock
\shownote{Accessed: 2026-03-11}.


\bibitem[Grossman et~al\mbox{.}(2023)]%
        {grossman2023compile}
\bibfield{author}{\bibinfo{person}{Aiden Grossman}, \bibinfo{person}{Ludger
  Paehler}, \bibinfo{person}{Konstantinos Parasyris}, \bibinfo{person}{Tal
  Ben-Nun}, \bibinfo{person}{Jacob Hegna}, \bibinfo{person}{William Moses},
  \bibinfo{person}{Jose M~Monsalve Diaz}, \bibinfo{person}{Mircea Trofin},
  {and} \bibinfo{person}{Johannes Doerfert}.} \bibinfo{year}{2023}\natexlab{}.
\newblock \showarticletitle{Compile: A large ir dataset from production
  sources}.
\newblock \bibinfo{journal}{\emph{arXiv preprint arXiv:2309.15432}}
  (\bibinfo{year}{2023}).
\newblock


\bibitem[Guo et~al\mbox{.}(2025)]%
        {guo2025deepseek}
\bibfield{author}{\bibinfo{person}{Daya Guo}, \bibinfo{person}{Dejian Yang},
  \bibinfo{person}{Haowei Zhang}, \bibinfo{person}{Junxiao Song},
  \bibinfo{person}{Ruoyu Zhang}, \bibinfo{person}{Runxin Xu},
  \bibinfo{person}{Qihao Zhu}, \bibinfo{person}{Shirong Ma},
  \bibinfo{person}{Peiyi Wang}, \bibinfo{person}{Xiao Bi}, {et~al\mbox{.}}}
  \bibinfo{year}{2025}\natexlab{}.
\newblock \showarticletitle{Deepseek-r1: Incentivizing reasoning capability in
  llms via reinforcement learning}.
\newblock \bibinfo{journal}{\emph{arXiv preprint arXiv:2501.12948}}
  (\bibinfo{year}{2025}).
\newblock


\bibitem[Guo and Moses(2022)]%
        {moses2022understanding}
\bibfield{author}{\bibinfo{person}{Zifan~(Carl) Guo} {and}
  \bibinfo{person}{William~S. Moses}.} \bibinfo{year}{2022}\natexlab{}.
\newblock \bibinfo{title}{{Understanding high-level properties of low-level
  programs through transformers}}.
\newblock
\newblock
\shownote{\url{https://api.semanticscholar.org/CorpusID:251439807}}.


\bibitem[Jiang et~al\mbox{.}(2025)]%
        {jiang2025can}
\bibfield{author}{\bibinfo{person}{Hailong Jiang}, \bibinfo{person}{Jianfeng
  Zhu}, \bibinfo{person}{Yao Wan}, \bibinfo{person}{Bo Fang},
  \bibinfo{person}{Hongyu Zhang}, \bibinfo{person}{Ruoming Jin}, {and}
  \bibinfo{person}{Qiang Guan}.} \bibinfo{year}{2025}\natexlab{}.
\newblock \showarticletitle{Can Large Language Models Understand Intermediate
  Representations in Compilers?}
\newblock \bibinfo{journal}{\emph{arXiv preprint arXiv:2502.06854}}
  (\bibinfo{year}{2025}).
\newblock


\bibitem[Kocetkov et~al\mbox{.}(2022)]%
        {kocetkov2022stack}
\bibfield{author}{\bibinfo{person}{Denis Kocetkov}, \bibinfo{person}{Raymond
  Li}, \bibinfo{person}{Loubna~Ben Allal}, \bibinfo{person}{Jia Li},
  \bibinfo{person}{Chenghao Mou}, \bibinfo{person}{Carlos~Mu{\~n}oz Ferrandis},
  \bibinfo{person}{Yacine Jernite}, \bibinfo{person}{Margaret Mitchell},
  \bibinfo{person}{Sean Hughes}, \bibinfo{person}{Thomas Wolf},
  {et~al\mbox{.}}} \bibinfo{year}{2022}\natexlab{}.
\newblock \showarticletitle{The stack: 3 tb of permissively licensed source
  code}.
\newblock \bibinfo{journal}{\emph{arXiv preprint arXiv:2211.15533}}
  (\bibinfo{year}{2022}).
\newblock


\bibitem[Lattner and Adve(2004)]%
        {lattner2004llvm}
\bibfield{author}{\bibinfo{person}{Chris Lattner} {and} \bibinfo{person}{Vikram
  Adve}.} \bibinfo{year}{2004}\natexlab{}.
\newblock \showarticletitle{LLVM: A compilation framework for lifelong program
  analysis \& transformation}. In \bibinfo{booktitle}{\emph{International
  symposium on code generation and optimization, 2004. CGO 2004.}} IEEE,
  \bibinfo{pages}{75--86}.
\newblock


\bibitem[Lattner et~al\mbox{.}(2021)]%
        {mlir}
\bibfield{author}{\bibinfo{person}{Chris Lattner}, \bibinfo{person}{Mehdi
  Amini}, \bibinfo{person}{Uday Bondhugula}, \bibinfo{person}{Albert Cohen},
  \bibinfo{person}{Andy Davis}, \bibinfo{person}{Jacques Pienaar},
  \bibinfo{person}{River Riddle}, \bibinfo{person}{Tatiana Shpeisman},
  \bibinfo{person}{Nicolas Vasilache}, {and} \bibinfo{person}{Oleksandr
  Zinenko}.} \bibinfo{year}{2021}\natexlab{}.
\newblock \showarticletitle{{MLIR}: Scaling Compiler Infrastructure for Domain
  Specific Computation}. In \bibinfo{booktitle}{\emph{2021 IEEE/ACM
  International Symposium on Code Generation and Optimization (CGO)}}.
  \bibinfo{pages}{2--14}.
\newblock
\href{https://doi.org/10.1109/CGO51591.2021.9370308}{doi:\nolinkurl{10.1109/CGO51591.2021.9370308}}


\bibitem[Li et~al\mbox{.}(2022)]%
        {li2022competition}
\bibfield{author}{\bibinfo{person}{Yujia Li}, \bibinfo{person}{David Choi},
  \bibinfo{person}{Junyoung Chung}, \bibinfo{person}{Nate Kushman},
  \bibinfo{person}{Julian Schrittwieser}, \bibinfo{person}{R{\'e}mi Leblond},
  \bibinfo{person}{Tom Eccles}, \bibinfo{person}{James Keeling},
  \bibinfo{person}{Felix Gimeno}, \bibinfo{person}{Agustin Dal~Lago},
  {et~al\mbox{.}}} \bibinfo{year}{2022}\natexlab{}.
\newblock \showarticletitle{{Competition-level code generation with
  Alphacode}}.
\newblock \bibinfo{journal}{\emph{Science}} \bibinfo{volume}{378},
  \bibinfo{number}{6624} (\bibinfo{year}{2022}), \bibinfo{pages}{1092--1097}.
\newblock


\bibitem[LLVM(2004)]%
        {llvm04frontend}
\bibfield{author}{\bibinfo{person}{LLVM}.} \bibinfo{year}{2004}\natexlab{}.
\newblock \bibinfo{title}{{llvm-gcc: LLVM C front-end}}.
\newblock
\newblock
\shownote{\url{https://releases.llvm.org/1.3/docs/CommandGuide/html/llvmgcc.html}}.


\bibitem[{LLVM Project}(2024)]%
        {llvm_langref}
\bibfield{author}{\bibinfo{person}{{LLVM Project}}.}
  \bibinfo{year}{2024}\natexlab{}.
\newblock \bibinfo{title}{LLVM Language Reference Manual}.
\newblock \bibinfo{howpublished}{\url{https://llvm.org/docs/LangRef.html}}.
\newblock
\newblock
\shownote{Accessed: 2026-03-16}.


\bibitem[Lopes et~al\mbox{.}(2021)]%
        {lopes2021alive2}
\bibfield{author}{\bibinfo{person}{Nuno~P Lopes}, \bibinfo{person}{Juneyoung
  Lee}, \bibinfo{person}{Chung-Kil Hur}, \bibinfo{person}{Zhengyang Liu}, {and}
  \bibinfo{person}{John Regehr}.} \bibinfo{year}{2021}\natexlab{}.
\newblock \showarticletitle{Alive2: bounded translation validation for LLVM}.
  In \bibinfo{booktitle}{\emph{Proceedings of the 42nd ACM SIGPLAN
  International Conference on Programming Language Design and Implementation}}.
  \bibinfo{pages}{65--79}.
\newblock


\bibitem[Loshchilov and Hutter(2017)]%
        {loshchilov2017decoupled}
\bibfield{author}{\bibinfo{person}{Ilya Loshchilov} {and}
  \bibinfo{person}{Frank Hutter}.} \bibinfo{year}{2017}\natexlab{}.
\newblock \showarticletitle{Decoupled weight decay regularization}.
\newblock \bibinfo{journal}{\emph{arXiv preprint arXiv:1711.05101}}
  (\bibinfo{year}{2017}).
\newblock


\bibitem[Mu(2024)]%
        {mu2024mlirgccjit}
\bibfield{author}{\bibinfo{person}{Sirui Mu}.} \bibinfo{year}{2024}\natexlab{}.
\newblock \bibinfo{title}{{mlir-gccjit}}.
\newblock
\newblock
\shownote{\url{https://github.com/Lancern/mlir-gccjit}}.


\bibitem[Nam et~al\mbox{.}(2024)]%
        {nam2024using}
\bibfield{author}{\bibinfo{person}{Daye Nam}, \bibinfo{person}{Andrew Macvean},
  \bibinfo{person}{Vincent Hellendoorn}, \bibinfo{person}{Bogdan Vasilescu},
  {and} \bibinfo{person}{Brad Myers}.} \bibinfo{year}{2024}\natexlab{}.
\newblock \showarticletitle{{Using an LLM to help with code understanding}}. In
  \bibinfo{booktitle}{\emph{46th International Conference on Software
  Engineering}}. \bibinfo{pages}{1--13}.
\newblock


\bibitem[OpenAI(2025)]%
        {openai25codex}
\bibfield{author}{\bibinfo{person}{OpenAI}.} \bibinfo{year}{2025}\natexlab{}.
\newblock \bibinfo{title}{Codex}.
\newblock
\newblock
\shownote{\url{https://openai.com/codex}}.


\bibitem[Pan et~al\mbox{.}(2025)]%
        {pan2025code}
\bibfield{author}{\bibinfo{person}{Zhenyu Pan}, \bibinfo{person}{Xuefeng Song},
  \bibinfo{person}{Yunkun Wang}, \bibinfo{person}{Rongyu Cao},
  \bibinfo{person}{Binhua Li}, \bibinfo{person}{Yongbin Li}, {and}
  \bibinfo{person}{Han Liu}.} \bibinfo{year}{2025}\natexlab{}.
\newblock \showarticletitle{{Do Code LLMs Understand Design Patterns?}}. In
  \bibinfo{booktitle}{\emph{IEEE/ACM International Workshop on Large Language
  Models for Code (LLM4Code)}}. IEEE, \bibinfo{pages}{209--212}.
\newblock


\bibitem[Paul et~al\mbox{.}(2024)]%
        {paul2024ircoder}
\bibfield{author}{\bibinfo{person}{Indraneil Paul}, \bibinfo{person}{Goran
  Glava{\v{s}}}, {and} \bibinfo{person}{Iryna Gurevych}.}
  \bibinfo{year}{2024}\natexlab{}.
\newblock \showarticletitle{Ircoder: Intermediate representations make language
  models robust multilingual code generators}.
\newblock \bibinfo{journal}{\emph{arXiv preprint arXiv:2403.03894}}
  (\bibinfo{year}{2024}).
\newblock


\bibitem[Penedo et~al\mbox{.}(2025)]%
        {penedo2025codeforces}
\bibfield{author}{\bibinfo{person}{Guilherme Penedo}, \bibinfo{person}{Anton
  Lozhkov}, \bibinfo{person}{Hynek Kydlíček}, \bibinfo{person}{Loubna~Ben
  Allal}, \bibinfo{person}{Edward Beeching}, \bibinfo{person}{Agustín~Piqueres
  Lajarín}, \bibinfo{person}{Quentin Gallouédec}, \bibinfo{person}{Nathan
  Habib}, \bibinfo{person}{Lewis Tunstall}, {and} \bibinfo{person}{Leandro von
  Werra}.} \bibinfo{year}{2025}\natexlab{}.
\newblock \bibinfo{title}{CodeForces}.
\newblock
  \bibinfo{howpublished}{\url{https://huggingface.co/datasets/open-r1/codeforces}}.
\newblock


\bibitem[Project(2025)]%
        {gccrs}
\bibfield{author}{\bibinfo{person}{Rust-GCC Project}.}
  \bibinfo{year}{2025}\natexlab{}.
\newblock \bibinfo{title}{{gccrs}: GCC Rust Front-End}.
\newblock \bibinfo{howpublished}{\url{https://github.com/Rust-GCC/gccrs}}.
\newblock
\newblock
\shownote{Accessed: 11 December 2025}.


\bibitem[Ranasinghe et~al\mbox{.}(2025)]%
        {ranasinghe2025llm}
\bibfield{author}{\bibinfo{person}{Nishath~Rajiv Ranasinghe},
  \bibinfo{person}{Shawn~M Jones}, \bibinfo{person}{Michal Kucer},
  \bibinfo{person}{Ayan Biswas}, \bibinfo{person}{Daniel O’Malley},
  \bibinfo{person}{Alexander Most}, \bibinfo{person}{Selma~Liliane Wanna},
  {and} \bibinfo{person}{Ajay Sreekumar}.} \bibinfo{year}{2025}\natexlab{}.
\newblock \showarticletitle{{LLM-assisted translation of legacy FORTRAN codes
  to C++: A cross-platform study}}. In \bibinfo{booktitle}{\emph{1st Workshop
  on AI and Scientific Discovery: Directions and Opportunities}}.
  \bibinfo{pages}{58--69}.
\newblock


\bibitem[Rifkin(2024)]%
        {rifkin2024wyrm}
\bibfield{author}{\bibinfo{person}{Jeremy Rifkin}.}
  \bibinfo{year}{2024}\natexlab{}.
\newblock \bibinfo{title}{{Wyrm}}.
\newblock
\newblock
\shownote{\url{https://github.com/jeremy-rifkin/wyrm}}.


\bibitem[Roziere et~al\mbox{.}(2020)]%
        {roziere2020unsupervised}
\bibfield{author}{\bibinfo{person}{Baptiste Roziere},
  \bibinfo{person}{Marie-Anne Lachaux}, \bibinfo{person}{Lowik Chanussot},
  {and} \bibinfo{person}{Guillaume Lample}.} \bibinfo{year}{2020}\natexlab{}.
\newblock \showarticletitle{{Unsupervised translation of programming
  languages}}.
\newblock \bibinfo{journal}{\emph{Advances in neural information processing
  systems}}  \bibinfo{volume}{33} (\bibinfo{year}{2020}),
  \bibinfo{pages}{20601--20611}.
\newblock


\bibitem[Su et~al\mbox{.}(2024)]%
        {su2024roformer}
\bibfield{author}{\bibinfo{person}{Jianlin Su}, \bibinfo{person}{Murtadha
  Ahmed}, \bibinfo{person}{Yu Lu}, \bibinfo{person}{Shengfeng Pan},
  \bibinfo{person}{Wen Bo}, {and} \bibinfo{person}{Yunfeng Liu}.}
  \bibinfo{year}{2024}\natexlab{}.
\newblock \showarticletitle{Roformer: Enhanced transformer with rotary position
  embedding}.
\newblock \bibinfo{journal}{\emph{Neurocomputing}}  \bibinfo{volume}{568}
  (\bibinfo{year}{2024}), \bibinfo{pages}{127063}.
\newblock


\bibitem[Szafraniec et~al\mbox{.}(2022)]%
        {szafraniec2022code}
\bibfield{author}{\bibinfo{person}{Marc Szafraniec}, \bibinfo{person}{Baptiste
  Roziere}, \bibinfo{person}{Hugh Leather}, \bibinfo{person}{Francois Charton},
  \bibinfo{person}{Patrick Labatut}, {and} \bibinfo{person}{Gabriel Synnaeve}.}
  \bibinfo{year}{2022}\natexlab{}.
\newblock \showarticletitle{{Code translation with compiler representations}}.
\newblock \bibinfo{journal}{\emph{arXiv preprint arXiv:2207.03578}}
  (\bibinfo{year}{2022}).
\newblock


\bibitem[Tan et~al\mbox{.}(2023)]%
        {tan2023splendid}
\bibfield{author}{\bibinfo{person}{Zujun Tan}, \bibinfo{person}{Yebin Chon},
  \bibinfo{person}{Michael Kruse}, \bibinfo{person}{Johannes Doerfert},
  \bibinfo{person}{Ziyang Xu}, \bibinfo{person}{Brian Homerding},
  \bibinfo{person}{Simone Campanoni}, {and} \bibinfo{person}{David~I August}.}
  \bibinfo{year}{2023}\natexlab{}.
\newblock \showarticletitle{{Splendid: Supporting parallel LLVM-IR enhanced
  natural decompilation for interactive development}}. In
  \bibinfo{booktitle}{\emph{Proceedings of the 28th ACM International
  Conference on Architectural Support for Programming Languages and Operating
  Systems, Volume 3}}. \bibinfo{pages}{679--693}.
\newblock


\bibitem[Team et~al\mbox{.}(2025)]%
        {team2025kimi}
\bibfield{author}{\bibinfo{person}{Kimi Team}, \bibinfo{person}{Yifan Bai},
  \bibinfo{person}{Yiping Bao}, \bibinfo{person}{Guanduo Chen},
  \bibinfo{person}{Jiahao Chen}, \bibinfo{person}{Ningxin Chen},
  \bibinfo{person}{Ruijue Chen}, \bibinfo{person}{Yanru Chen},
  \bibinfo{person}{Yuankun Chen}, \bibinfo{person}{Yutian Chen},
  {et~al\mbox{.}}} \bibinfo{year}{2025}\natexlab{}.
\newblock \showarticletitle{Kimi k2: Open agentic intelligence}.
\newblock \bibinfo{journal}{\emph{arXiv preprint arXiv:2507.20534}}
  (\bibinfo{year}{2025}).
\newblock


\bibitem[Toor(2022)]%
        {toor2022decompilation}
\bibfield{author}{\bibinfo{person}{Tejvinder Toor}.}
  \bibinfo{year}{2022}\natexlab{}.
\newblock \emph{\bibinfo{title}{{Decompilation of Binaries into LLVM IR for
  Automated Analysis}}}.
\newblock \bibinfo{thesistype}{Ph.\,D. Dissertation}.
  \bibinfo{school}{University of Waterloo}.
\newblock


\bibitem[Valenzuela et~al\mbox{.}(2025)]%
        {valenzuela2025from}
\bibfield{author}{\bibinfo{person}{Andrea Valenzuela}, \bibinfo{person}{Marta
  Gonzalez-Mallo}, \bibinfo{person}{Cristian Gutierrez}, \bibinfo{person}{Dario
  Garcia-Gasulla}, \bibinfo{person}{Gokcen Kestor}, {and} \bibinfo{person}{Sara
  Royuela}.} \bibinfo{year}{2025}\natexlab{}.
\newblock \showarticletitle{{From C to Rust: Evaluating LLM Capabilities in
  Transpilation Through Compilation Errors}}. In
  \bibinfo{booktitle}{\emph{International Conference on High Performance
  Computing}}. Springer, \bibinfo{pages}{311--324}.
\newblock


\bibitem[Vaswani et~al\mbox{.}(2017)]%
        {vaswani2017attention}
\bibfield{author}{\bibinfo{person}{Ashish Vaswani}, \bibinfo{person}{Noam
  Shazeer}, \bibinfo{person}{Niki Parmar}, \bibinfo{person}{Jakob Uszkoreit},
  \bibinfo{person}{Llion Jones}, \bibinfo{person}{Aidan~N Gomez},
  \bibinfo{person}{{\L}ukasz Kaiser}, {and} \bibinfo{person}{Illia
  Polosukhin}.} \bibinfo{year}{2017}\natexlab{}.
\newblock \showarticletitle{{Attention is all you need}}.
\newblock \bibinfo{journal}{\emph{Advances in neural information processing
  systems}}  \bibinfo{volume}{30} (\bibinfo{year}{2017}).
\newblock


\bibitem[Wirth(1982)]%
        {Wirth1982Modula2}
\bibfield{author}{\bibinfo{person}{Niklaus Wirth}.}
  \bibinfo{year}{1982}\natexlab{}.
\newblock \bibinfo{booktitle}{\emph{Programming in Modula-2}}.
\newblock \bibinfo{publisher}{Springer-Verlag}, \bibinfo{address}{Berlin,
  Heidelberg}.
\newblock


\bibitem[Yang et~al\mbox{.}(2025)]%
        {yang2025qwen3}
\bibfield{author}{\bibinfo{person}{An Yang}, \bibinfo{person}{Anfeng Li},
  \bibinfo{person}{Baosong Yang}, \bibinfo{person}{Beichen Zhang},
  \bibinfo{person}{Binyuan Hui}, \bibinfo{person}{Bo Zheng},
  \bibinfo{person}{Bowen Yu}, \bibinfo{person}{Chang Gao},
  \bibinfo{person}{Chengen Huang}, \bibinfo{person}{Chenxu Lv},
  {et~al\mbox{.}}} \bibinfo{year}{2025}\natexlab{}.
\newblock \showarticletitle{Qwen3 technical report}.
\newblock \bibinfo{journal}{\emph{arXiv preprint arXiv:2505.09388}}
  (\bibinfo{year}{2025}).
\newblock


\bibitem[Yeti{\c{s}}tiren et~al\mbox{.}(2023)]%
        {yeticstiren2023evaluating}
\bibfield{author}{\bibinfo{person}{Burak Yeti{\c{s}}tiren},
  \bibinfo{person}{I{\c{s}}{\i}k {\"O}zsoy}, \bibinfo{person}{Miray Ayerdem},
  {and} \bibinfo{person}{Eray T{\"u}z{\"u}n}.} \bibinfo{year}{2023}\natexlab{}.
\newblock \showarticletitle{{Evaluating the code quality of AI-assisted code
  generation tools: An empirical study on GitHub Copilot, Amazon CodeWhisperer,
  and ChatGPT}}.
\newblock \bibinfo{journal}{\emph{arXiv preprint arXiv:2304.10778}}
  (\bibinfo{year}{2023}).
\newblock


\bibitem[Zhang et~al\mbox{.}(2026)]%
        {zhang2026new}
\bibfield{author}{\bibinfo{person}{Shuoming Zhang}, \bibinfo{person}{Jiacheng
  Zhao}, \bibinfo{person}{Qiuchu Yu}, \bibinfo{person}{Chunwei Xia},
  \bibinfo{person}{Zheng Wang}, \bibinfo{person}{Xiaobing Feng}, {and}
  \bibinfo{person}{Huimin Cui}.} \bibinfo{year}{2026}\natexlab{}.
\newblock \showarticletitle{The new compiler stack: a survey on the synergy of
  LLMs and compilers}.
\newblock \bibinfo{journal}{\emph{CCF Transactions on High Performance
  Computing}} (\bibinfo{year}{2026}), \bibinfo{pages}{1--32}.
\newblock


\end{thebibliography}

\end{document}